\documentclass[a4paper,fleqn,usenatbib]{mnras}

\usepackage{newtxtext,newtxmath}

\usepackage[T1]{fontenc}
\usepackage{ae,aecompl}

\usepackage{graphicx}	%
\usepackage{amsmath}	%
\usepackage{multirow}
\usepackage{pdflscape}
\usepackage[normalem]{ulem}
\usepackage{xcolor}

\newcommand{\galex}{\textit{GALEX}}

\newcommand{\um}{$\mu\mathrm{m}$}
\newcommand{\hone}{\ion{H}{I}}
\newcommand{\htwo}{H$_2$}

\title[Gas \& dust in transition galaxies]{The cold gas and dust properties of red star-forming galaxies}

\author[R. Chown et al.]{Ryan Chown,$^{1,2}$\thanks{E-mail: ryan.chown@mail.mcgill.ca}
Laura Parker,$^{2}$
Christine D. Wilson,$^{2}$ 
Toby Brown,$^{2,3}$ 
Fraser Evans,$^{4}$
\newauthor 
Yang Gao,$^{5}$ 
Ho Seong Hwang,$^{6,7}$ 
Lihwai Lin,$^{8}$
Amelie Saintonge,$^{9}$
Mark Sargent,$^{10,11}$
\newauthor 
Matthew Smith$^{12}$
and Ting Xiao$^{13}$
\\
$^{1}$Department of Physics \& Astronomy, University of Western Ontario, London, ON N6A 3K7, Canada \\
$^{2}$Department of Physics \& Astronomy, McMaster University, Hamilton, ON L8S 4M1, Canada \\
$^{3}$Herzberg Astronomy and Astrophysics Research Centre, National Research Council of Canada, 5071 West Saanich Rd., Victoria, BC V9E 2E7, Canada\\
$^{4}$Leiden Observatory, Leiden University, PO Box 9513, NL-2300 RA Leiden, the Netherlands \\
$^{5}$Purple Mountain Observatory \& Key Lab of Radio Astronomy, Chinese Academy of Sciences, Nanjing 210034, China \\
$^{6}$Astronomy Program, Department of Physics and Astronomy, Seoul National University, 1 Gwanak-ro, Gwanak-gu, Seoul 08826, Korea \\
$^{7}$SNU Astronomy Research Center, Seoul National University, 1 Gwanak-ro, Gwanak-gu, Seoul 08826, Korea\\
$^{8}$Institute of Astronomy \& Astrophysics, Academia Sinica, Taipei 10617, Taiwan \\
$^{9}$Department of Physics and Astronomy, University College London, Gower Street, London, WC1E 6BT, UK\\
$^{10}$Astronomy Centre, Department of Physics \& Astronomy, University of Sussex, Brighton, BN1 9QH, England\\
$^{11}$International Space Science Institute (ISSI), Hallerstrasse 6, 3012 Bern, Switzerland\\
$^{12}$School of Physics \& Astronomy, Cardiff University, The Parade, Cardiff CF24 3AA, UK \\
$^{13}$Department of Physics, Zhejiang University, Hangzhou, Zhejiang 310027, China
}

\date{Accepted XXX. Received YYY; in original form ZZZ}

\pubyear{2022}

\begin{document}
\label{firstpage}
\pagerange{\pageref{firstpage}--\pageref{lastpage}}
\maketitle

\begin{abstract}

We study the cold gas and dust properties for a sample of red star forming galaxies called ``red misfits.''
We collect single-dish CO observations and \hone\ observations from representative samples of low-redshift galaxies, as well as our own JCMT CO observations of red misfits. We also obtain SCUBA-2 850 \micron\ 
observations for a subset of these galaxies.
With these data we compare the molecular gas, total cold gas, and dust properties of red misfits against those of their blue counterparts (``blue actives'') taking non-detections into account using a survival analysis technique. We compare these properties at fixed position in the $\log \mathrm{SFR}$-$\log M_\star$ plane, as well as versus offset from the star-forming main sequence.
Compared to blue actives, red misfits have slightly longer molecular gas depletion times, similar total
gas depletion times, significantly lower molecular- and total-gas mass fractions, lower dust-to-stellar mass ratios, similar dust-to-gas ratios, and a significantly flatter slope in the $\log M_\mathrm{mol}$-$\log M_\star$ plane.
Our results suggest that red misfits as a population are likely quenching due to a shortage in gas supply.

\end{abstract}


\section{Introduction}

A key finding from large surveys of the local universe such as the Sloan Digital Sky Survey \citep[SDSS;][]{york2000}
is that
the vast majority of galaxies in the nearby universe tend to fall into one of two categories: 
a star-forming ``main sequence'' (SFMS) where star formation rate (SFR) and stellar mass $M_\star$ are 
well-correlated, or the quiescent population where SFRs are low and not well-correlated with $M_\star$.
In colour-magnitude space, star-forming galaxies are found in the diffusely-populated region called the ``blue cloud,'' while quiescent galaxies have red colours and form a tight correlation between colour and magnitude
called the ``red sequence.'' 
A small but significant number of galaxies lie in the so-called ``green valley'' between the main sequence and 
red cloud \citep[e.g][]{salim2014}.

Studying the relationship between galaxy properties and 
galaxy position in the SFR-$M_\star$ plane 
has provided insight into 
which physical processes are responsible for evolution in this plane \citep{saintonge2022}. 
One approach to tackle this question is to focus on populations
with intermediate specific SFR (SSFR $\equiv$ SFR/$M_\star$), which may be evolving away from or toward the SFMS \citep[e.g.][]{salim2014, schawinski2014, smethurst2015, salim2018, li2015, belfiore2017, lin2017, eales2018, coenda2018, mancini2019, lin2022, brownson2020}. 
However, some works argue that the green valley exists due to 
 observational biases rather than physical processes
\citep{schawinski2014, eales2018}.

According to the gas-regulator model \citep{lilly2013}, processes which affect inflows, outflows, 
and consumption of 
gas determine the star formation rate of a galaxy.
Gas depletion time $t_\mathrm{gas}$
\begin{equation}
t_\mathrm{gas} \> \mathrm{[yr]} \equiv \frac{M_\mathrm{gas} \> \mathrm{[M_\odot]}}{\mathrm{SFR \> [M_\odot\> yr^{-1}}]}
\end{equation}
is the time it would take for a gas reservoir to turn into stars assuming  
none of this gas dissipates, no new gas is accreted,
gas is not returned to the interstellar medium (ISM) via stellar evolution, and that the SFR is constant over time.
Although these assumptions are not physically realistic, 
it is useful to think of $t_\mathrm{gas}$ as a proxy for 
the efficiency with which gas is converted into stars.
In the literature, the reciprocal of gas depletion time is often referred to as 
``star formation efficiency''  \citep[SFE, e.g.][]{leroy2008, saintonge2017}. To avoid confusion
with the theoretical star formation efficiency $\epsilon_\mathrm{SF}$, namely 
the fraction of a gas reservoir that forms stars before it dissipates, or the more commonly-used efficiency per free fall time, 
in this work we will write $t_\mathrm{gas}^{-1}$ instead of ``SFE.''

Observations of the total cold atomic and molecular gas reservoirs in large samples of nearby galaxies
such as the Galaxy Evolution EXplorer (\galex) Arecibo SDSS Survey \citep[xGASS;][]{catinella2018},
the CO Legacy Database for GASS \citep[xCOLD GASS;][]{saintonge2011, saintonge2017}, 
and the James Clerk Maxwell Telescope (JCMT) dust and gas In Nearby Galaxies Legacy Exploration \citep[JINGLE;][]{saintonge2018}, 
have found 
that $t_\mathrm{gas}$ is correlated with offset from the SFMS 
such that $t_\mathrm{gas}$ decreases with increasing offset from the main sequence 
\begin{equation}\label{eq:dms}
\Delta \mathrm{MS} \> \mathrm{[dex]} \equiv \log \frac{\mathrm{SFR}}{\mathrm{SFR_{MS}}},
\end{equation}
where $\mathrm{SFR_{MS}}$ (SFR as a function of stellar mass) defines the SFMS.
\citet{tacconi2018} find that the trend between $t_\mathrm{gas}$ and $\Delta \mathrm{MS}$ persists from $z = 4$ to 0.
\citet{colombo2020} find that declining molecular gas mass fractions drive galaxies off of the SFMS, and that once
a galaxy is quenched, $t_\mathrm{gas}$ is more important than molecular gas mass
 in determining the SFR.
With the ALMA-MaNGA QUEnching and STar formation (ALMaQUEST) sample, \citet{lin2020a} and \citet{ellison2020} find that local variations in $t_\mathrm{gas}$ cause regions to depart from the spatially-resolved SFMS (that is, the SFMS based on SFR and $M_\star$ surface densities in sub-regions of galaxies rather than galaxy-integrated measurements). 
\citet{brownson2020} study seven green valley galaxies, and find that $t_\mathrm{gas}$ and $f_\mathrm{gas}$ (gas mass divided by stellar mass) are equally important in driving departures from the SFMS.
A recent analysis of xCOLD GASS and xGASS data \citep{feldmann2020} found that after accounting for galaxy selection biases (e.g. stellar mass, SFR) and observational uncertainties the correlation between $\log t_\mathrm{mol}$ (the depletion time of molecular gas only) and  $\Delta \mathrm{MS}$ flattens significantly, from 
$\log t_\mathrm{mol} \propto -0.5~\Delta \mathrm{MS}$ to $\log t_\mathrm{mol} \propto -0.24~\Delta \mathrm{MS}$. In other words, they find that
$t_\mathrm{mol}$ has a small but significant dependence on offset from the SFMS after accounting for selection effects and observational uncertainties. 
This nearly flat relationship between $\log t_\mathrm{mol}$ and $\Delta \mathrm{MS}$ echoes the findings of \citet{sargent2014}.

A major motivation of the present work is to improve our understanding of galaxy evolution
by focusing on the gas and dust properties of a galaxy population (red misfits), that is selected differently from the green valley and shows differences from that population, but 
whose star formation, similar to the green valley, is possibly in the act of quenching. 
Another motivation is to use a large multi-wavelength sample to compare the gas and dust properties of red misfits with the overall population of low-redshift star-forming galaxies.

We investigate a population of galaxies selected from SDSS to be optically red 
and actively forming stars \citep{evans2018}. This population, called ``red misfits,''
appears to have no preference for environment, has an elevated fraction of active galactic nuclei (AGN), and
accounts for about 10 per cent of low redshift galaxies across stellar masses from $\log M_\star = 9.5$ to $11.5$ \citep{evans2018}.
\citet{evans2018} compared the properties of red misfits with green valley galaxies; they find 
that about 30 per cent of red misfits also lie in the green valley. Although there are similarities in these populations (e.g. both have significant AGN fractions, both are dominated by intermediate morphologies,
and star formation is likely slowing down in both populations), \citet{evans2018} find
several differences. 
Unlike green valley galaxies, red misfits are not simply in between blue star forming and red dead galaxies in the $\log \mathrm{SFR}$-$\log M_\star$ plane.
Green valley galaxies with late-type morphologies 
are rarely found in haloes with masses larger than $10^{12}\> h^{-1} \>$M$_\odot$ \citep{schawinski2014}; while
red misfits are found in roughly 
the same proportions for all halo masses \citep{evans2018}.
Green valley galaxies lie between the blue star-forming, and red-and-dead populations, suggesting that they represent
an intermediate stage of galaxy evolution \citep{salim2018},
while red misfits, in contrast, lie below, on, and above the SFMS while being red in colour, making their 
average stage of evolution less obvious and making them an interesting population to explore further \citep{evans2018}.

The primary goal of the present work 
is to better understand the evolutionary state of red misfit galaxies by studying their cold gas and dust properties.
We use a combination of two of the largest samples of 
CO in the local universe (xCOLD GASS and JINGLE), and sub-millimeter observations of a large number of them (from JINGLE and our own observations), as well as \hone\ measurements from xGASS and the Arecibo Legacy Fast ALFA (ALFALFA) $\alpha.100$ catalog \citep{haynes2018}
to compare the interstellar medium in red misfits with their blue counterparts (``blue actives'') and to try to understand the nature of red misfits.

We assume a flat $\Lambda$CDM cosmology with $H_0=70~\mathrm{km~s^{-1}~Mpc^{-1}}$, $\Omega_\mathrm{m,0}=0.27$, and $T_\mathrm{CMB,0}=2.275$ K.

\section{Data and Data Processing}\label{sec:data}

\subsection{Star formation rates, stellar masses, and other basic properties}\label{sec:gswlc}

\begin{figure*}
	\includegraphics[width=0.5\linewidth]{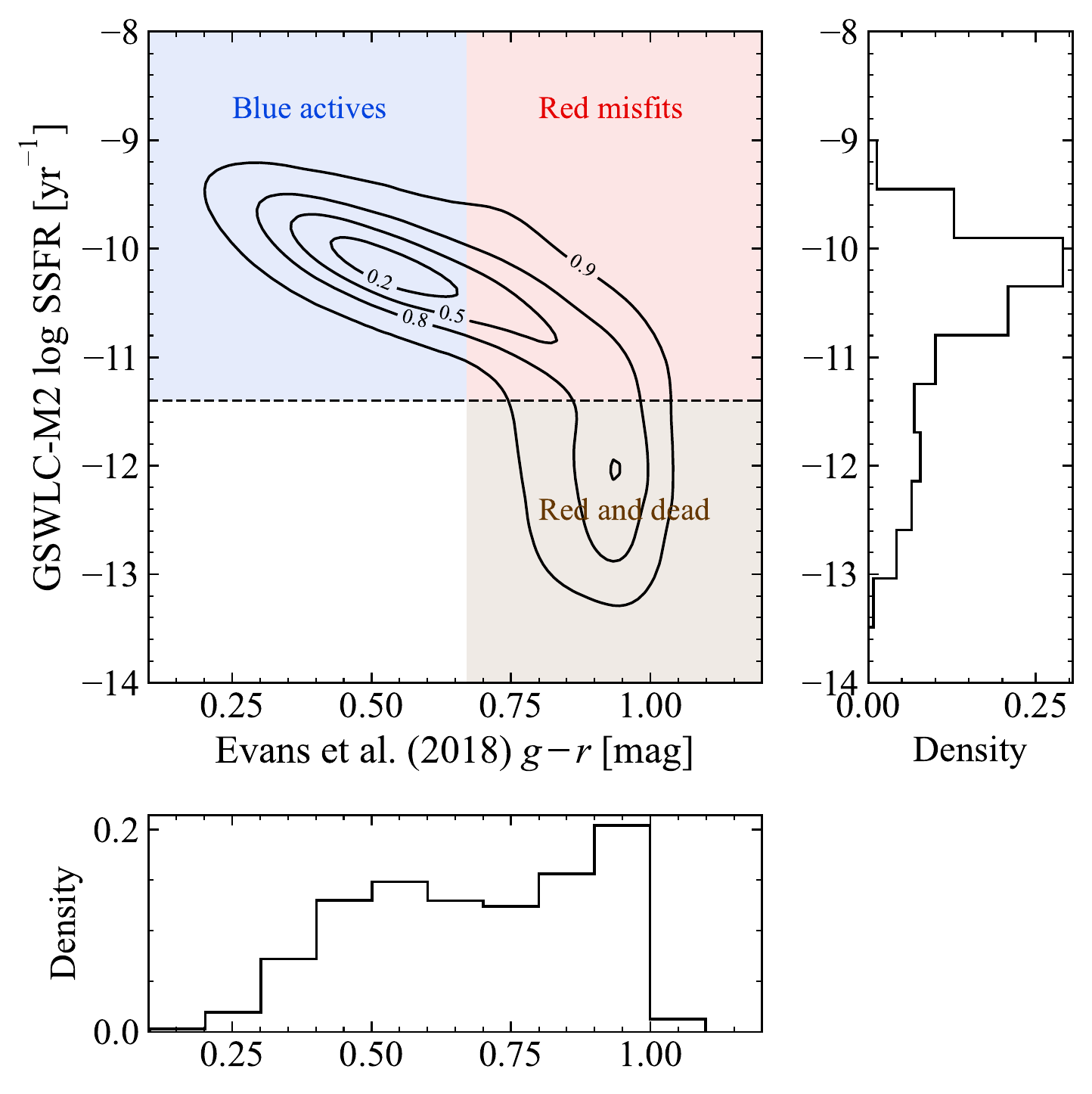}\includegraphics[width=0.5\linewidth]{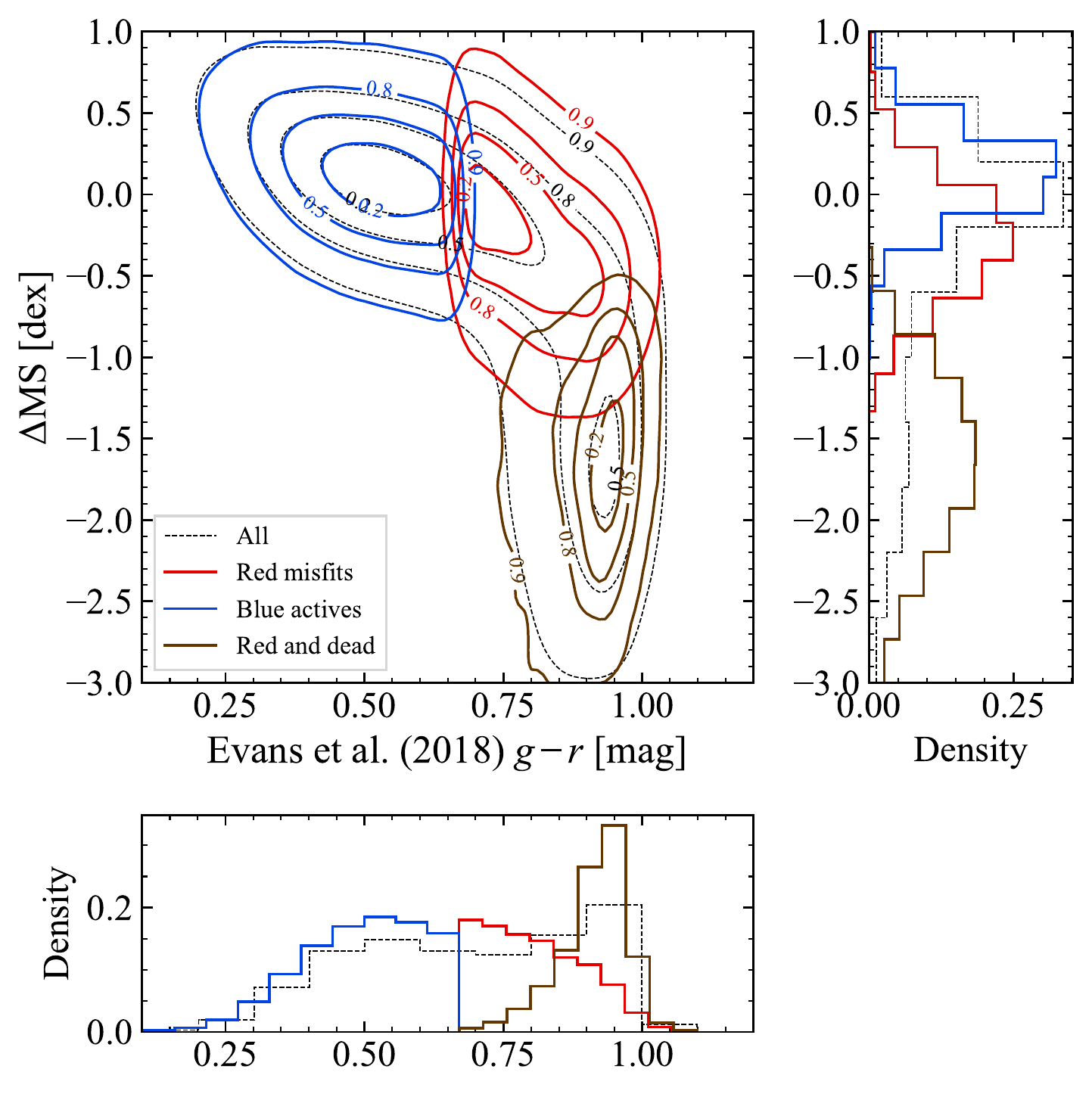}
	\caption[SSFR vs. $g-r$ colour of local galaxies]{\textit{Left}: SSFR vs. $g-r$ colour of all 118,769 galaxies in the intersection of the \citet{evans2018} and GSWLC-M2 catalogs. We used the GSWLC-M2 star formation rates and stellar masses to compute the dividing line between star-forming and passive galaxies (the horizontal line at $\log \> \mathrm{SSFR\> [yr^{-1}]} = -11.3$). This cut was determined by fitting the $\log \> \mathrm{SSFR}$ histogram with a double-Gaussian and calculating where the two Gaussians intersect. Red misfits are defined as galaxies that are star forming (above the horizontal line) and red in colour \citep[$g-r \geq 0.67$;][]{evans2018}. Red, passively evolving (``red and dead'') galaxies are also indicated. \textit{Right:} Offset from the SFMS vs. $g-r$ colour, for all galaxies (black dashed), blue actives (blue), red misfits (red), and red and dead galaxies (brown). Red misfits clearly occupy a wide range in $\Delta~\mathrm{MS}$ that is systematically higher than red and dead galaxies.
    }
    \label{fig:ssfr_gr_all}
\end{figure*}

Optical colours and specific star formation rates are required to select red misfits. Star formation
rates and stellar masses are also required to compute gas and dust-based quantities such as gas depletion times.
These optical data are taken from the following sources:
\begin{enumerate}
\item SDSS $g-r$ colours that have been extinction-corrected and inclination-corrected, taken from \citet{evans2018}. Some galaxies with CO measurements in our sample were not included in \citet{evans2018}. For this subset we computed $g-r$ colours using the same method.
\item SFR and stellar masses from UV + optical spectral energy distribution (SED) fitting taken from the GSWLC-M2 catalog \citep{salim2018}. The medium-depth (M2) measurements are ideal for star-forming galaxies, which are the focus of this work. Where available we use the M2 catalog, and for a small subset of galaxies we use the A2 catalog.
\end{enumerate}
The distribution of all 118,769 galaxies in the intersection of the \citet{evans2018} catalog and GSWLC-M2, in the log SSFR vs. $g-r$ colour plane ($0.01 \leq z \leq 0.12$, $9.0 \leq \log~M_\star \leq 11.9$) is shown in the left panel of Figure~\ref{fig:ssfr_gr_all}. We used the GSWLC-M2 star formation rates and stellar masses to compute the dividing line between star-forming and passive galaxies (the horizontal line at $\log \> \mathrm{SSFR\> [yr^{-1}]} = -11.3$). This cut was determined by fitting the $\log \> \mathrm{SSFR}$ histogram with a double-Gaussian and calculating where the two Gaussians intersect. In this work we focus on star forming galaxies, namely red misfits (upper right quadrant of this figure) and ``blue actives'' (upper left quadrant). Red misfits are defined as galaxies that are star forming (above the horizontal line) and red in colour \citep[$g-r \geq 0.67$;][]{evans2018}.
The right panel of Figure~\ref{fig:ssfr_gr_all} shows the relationship between $\Delta\mathrm{MS}$  (equation~\ref{eq:dms}) and $g-r$ colour, indicating that red misfits occupy a wide range in $\Delta~\mathrm{MS}$. 
Red misfits have a broader colour distribution and are systematically bluer than red and dead galaxies.
Red misfits have 
a narrower $\Delta \mathrm{MS}$ distribution than red and dead galaxies, and a median $\Delta \mathrm{MS}$ that is 1.4 dex closer to the MS than the red and dead population (Table~\ref{tab:delta_ms_gr_stats}).

\begin{table*}
\centering
\caption{Statistics of $\Delta~\mathrm{MS}$ and $g-r$ colour measured from galaxies in the parent sample.}
\label{tab:delta_ms_gr_stats}
\begin{tabular}{lllll}
\hline
Population & \multicolumn{2}{c}{Median} & \multicolumn{2}{c}{Robust standard deviation}\\
(1) & \multicolumn{2}{c}{(2)} & \multicolumn{2}{c}{(3)}\\

\hline
 & $\Delta~\mathrm{MS}$ [dex] & $g-r$ [mag] & $\Delta~\mathrm{MS}$ [dex] & $\Delta~g-r$ [mag] \\
\hline
All galaxies & $-0.13$ & $0.70$ & $0.56$ & $0.27$ \\
Red misfits & $-0.27$ & $0.80$ & $0.37$ & $0.11$ \\
Blue actives & $0.13$ & $0.51$ & $0.25$ & $0.12$ \\
Red and dead & $-1.65$ & $0.92$ & $0.56$ & $0.05$ \\
\hline
\multicolumn{5}{l}{(3) Defined as $1.4826~\mathrm{MAD}$ \citep{astropy-collaboration2018}.}\\
\end{tabular}
\end{table*}

\subsection{Single-dish CO observations}\label{sec:co}

We use CO observations from the following three sources:
\begin{enumerate}
\item \textbf{JCMT CO(2-1) measurements from the JINGLE survey} \citep{saintonge2018}. JINGLE is 
a representative sample of galaxies ranging from just below the star forming main sequence to the starburst regime.
The entire JINGLE sample was observed with SCUBA-2, while a subset of about 75 galaxies were observed in CO(2-1).
The JCMT beam at the frequency of CO(2-1) is 20 arcsec \citep{saintonge2018}.
Molecular gas mass is related to CO(2-1) luminosity $L_\mathrm{CO(2-1)}$ by
\begin{equation}
M_\mathrm{mol} \> \mathrm{[M_\odot]}
= r_{21}^{-1} \alpha_\mathrm{CO} L_\mathrm{CO(2-1)} \> \mathrm{[K\> km \> s^{-1} \> pc^2]},
\end{equation}
where $\alpha_\mathrm{CO}$ is the CO-to-\htwo\ conversion factor \citep{bolatto2013} and $r_{21}$ is the ratio of CO(2-1) to CO(1-0) intensities.
Note that in this work we use the subscript ``mol'' to indicate total molecular gas (hydrogen and helium).
In normal star-forming regions $\alpha_\mathrm{CO}$ is often assumed to be 4.35 $\mathrm{M_\odot}\>(\mathrm{K\> km \> s^{-1} \> pc^2})^{-1}$ \citep{bolatto2013}, which includes the contribution from helium (a factor of 1.36). 
For CO(2-1) measurements, one must assume a value of $r_{21}$. Variations from $r_{21} \sim 0.6$ \citep{yajima2021} to $\sim 0.8$ \citep{saintonge2017} have been observed. We use the commonly-used value of 0.7. The JINGLE analysis assumed a ratio of $r_{21}=0.7$ and $\alpha_\mathrm{CO} = 4.35$ \citep{saintonge2018}.
\item \textbf{IRAM 30 m CO(2-1) and some CO(1-0) fluxes from the xCOLD GASS survey} \citep{saintonge2017}. xCOLD GASS is a representative sample of CO emission in nearby galaxies. These galaxies were primarily selected from the xGASS survey \citep{catinella2018}. The IRAM 30 m beam sizes at the frequencies of the CO(2-1) and CO(1-0) lines are 11 arcsec and 22 arcsec respectively \citep{saintonge2017}.
The molecular gas masses in the xCOLD GASS catalog were computed using a metallicity-dependent $\alpha_\mathrm{CO}$. To be consistent with the JINGLE catalog we recalculated these molecular gas masses using $\alpha_\mathrm{CO} = 4.35$.
\item \textbf{Our own JCMT CO(2-1) measurements of red misfits.} These galaxies are from the JINGLE sample that
 were not scheduled to be observed in CO(2-1), but had already been observed with SCUBA-2. 
These data were reduced and converted into molecular gas masses using the same approach as for JINGLE galaxies [C. Wilson, private communication]. These measurements do not appear elsewhere in the literature and are provided in Table~\ref{tab:jcmt_18a}.
\end{enumerate}
The number of galaxies with CO measurements, and the sources of these measurements are shown 
in the first row of Table~\ref{tab:sample}.

\begin{table*}
\centering
\caption[Numbers of galaxies with CO, \hone, and 850 \micron\ measurements]{Numbers of galaxies with CO, \hone, and 850 \micron\ measurements, and the sources of these measurements.}
\label{tab:sample}
\begin{tabular}{llcc}
\hline
Measurement & Source(s) & \# Galaxies  & \# Non-Detections\\
\hline
CO & - JINGLE & \multirow{3}{*}{427} & \multirow{3}{*}{61} \\
 & - xCOLD GASS & & \\
 & - Our own observations from JCMT &  & \\
 \hline
\hone\ 21 cm $^\dagger$ & - xGASS & \multirow{3}{*}{369} & \multirow{3}{*}{65} \\
 & - JINGLE/Arecibo & & \\
 & - ALFALFA $\alpha.100$ & & \\
\hline
Dust (850 \micron) & - JINGLE & \multirow{2}{*}{209} & \multirow{2}{*}{106} \\
 & - Our own observations from JCMT & & \\
\hline
\multicolumn{3}{l}{$^\dagger$ Only galaxies with both CO and \hone\ observations are used.}\\
\end{tabular}
\end{table*}

\subsection{\hone\ observations}\label{sec:hi}

In addition to the molecular gas supply, we are interested in measuring the total 
gas mass
\begin{equation}\label{eq:totalgas}
M_\mathrm{gas} = 1.36(M_\mathrm{H_2}+M_\mathrm{HI}),
\end{equation}
where $M_\mathrm{H_2}$ is the molecular hydrogen mass and $M_\mathrm{HI}$ is the neutral hydrogen mass.
Note that in this work, the subscript ``gas'' refers to the total molecular and atomic gas as shown in
Equation~\ref{eq:totalgas}.
All of our \hone\ measurements were made using the Arecibo telescope, which has a beam size of $\sim 3.5$ arcmin \citep{catinella2018}.
We collected \hone\ measurements from the following sources:
\begin{enumerate}
\item \textbf{The ALFALFA $\alpha.100$ catalog \citep{haynes2018}.} We cross-matched the JINGLE sample with this catalog, which provided \hone\ measurements for 99 galaxies from the JINGLE sample. 
\item \textbf{The xGASS representative sample \citep{catinella2018}.} This sample provides \hone\ measurements for most of the galaxies in the xCOLD GASS sample.
\item Observations of a subset of the JINGLE sample using the Arecibo telescope (obtained by private communication with M. Smith). This sample consists of 60 JINGLE galaxies which were not observed as part of the ALFALFA survey. 
\end{enumerate}
The number of galaxies with \hone\ measurements, and the sources of these measurements are shown 
in the second row of Table~\ref{tab:sample}.

\subsection{Dust masses from sub-millimeter observations}\label{sec:dust}

We use SCUBA-2 850 \um\ flux densities $S_{850\>\mu\mathrm{m}}$ to estimate the cold dust mass of galaxies in our sample. The SCUBA-2 beam size at 850 \um\ is 13 arcsec. These measurements are from the following sources:
\begin{enumerate}
\item \textbf{SCUBA-2 850 \um\ measurements from the JINGLE survey} \citet{smith2019}. These data are available at \url{http://www.star.ucl.ac.uk/JINGLE/data.html}. On that page is a catalog of far-infrared and sub-mm photometry, from which we obtained 850 \um\ flux measurements.
\item \textbf{Our own SCUBA-2 850 \um\ measurements of a sample of red misfits}. These galaxies were selected from the xCOLD GASS sample. xCOLD GASS does not overlap significantly with far-infrared surveys -- this was the primary motivation for obtaining SCUBA-2 measurements of these galaxies. We present our 850 \um\ measurements in Table~\ref{tab:scuba2_18b}.
These measurements were processed in the same way as in \citet{smith2019} 
except we did not correct for CO(3-2) emission, which contributes a small amount to the observed 850 \um\ emission.
Across the JINGLE sample, the mean CO(3-2) correction is 10.1 per cent of the predicted 850 \um\ flux density \citep{smith2019}.
\end{enumerate}
The number of galaxies with SCUBA-2 850 \um\ measurements, and the sources of these measurements are shown 
in the third row of Table~\ref{tab:sample}.

To convert 850 \um\ flux densities $S_{850\>\mu\mathrm{m}}$ into dust masses, we first consider
the relationship between specific flux and dust mass at wavelength $\lambda$ assuming it emits as a modified blackbody
\begin{equation}\label{eq:flambda}
F_\lambda \> \mathrm{[W\>m^{-2}\>m^{-1}]} = M_\mathrm{dust}d_L^{-2} \kappa_\lambda(\beta)  B_\lambda(T),
\end{equation}
where $M_\mathrm{dust}$ is the dust mass in kg, 
$d_L$ is luminosity distance in m,
$\kappa_\lambda$ is the dust opacity in m$^2$ kg$^{-1}$, 
and $B_\lambda(T)$ is the Planck function
\begin{equation}
B_\lambda(T) = \frac{2hc^2}{\lambda^5}\frac{1}{\exp(hc/k_BT\lambda)-1}.
\end{equation}
Following \citet{lamperti2019}, dust opacity is given by
\begin{equation}
\kappa_\lambda(\beta) \> \mathrm{[m^2\>kg^{-1}]} = \kappa_0 \left(\frac{\lambda_0}{\lambda}\right)^\beta,
\end{equation}
where $\kappa_0=5.1\times 10^{-2}$ m$^2$ kg$^{-1}$ at 500 \um\ \citep{clark2016}, $\lambda_0=500$ \um, and $\beta$ is the spectral index.

The 850 \um\ flux density $S_{850\>\mu\mathrm{m}}$ in units of Jy can be converted into units of specific intensity via
\begin{equation}\label{eq:f850}
F_{850\> \mu \mathrm{m}} = 10^{-26} c \lambda^{-2} S_{850\>\mu\mathrm{m}}.
\end{equation}
Finally we can rearrange Equation~\ref{eq:flambda} for $M_\mathrm{dust}$ which gives
\begin{equation}\label{eq:mdust}
M_\mathrm{dust} = \frac{10^{-26} c \lambda^{-2} d_L^2 S_{850\>\mu\mathrm{m}}}{\kappa_\lambda(\beta)B_\lambda(T)}.
\end{equation}
We use the scaling relation for $\beta$ from Equation 35 in \citet{lamperti2019}
\begin{equation}\label{eq:lamperti_beta}
\beta = a_1 \log M_\star + a_2 \log A + a_3 \log [12+\log(\mathrm{O/H})] + a_4,
\end{equation}
where $A=\pi r_{50}^2$ in kpc$^2$ is the surface area corresponding to the SDSS $i$-band half-light radius $r_{50}$,
$12+\log(\mathrm{O/H})$ is the gas-phase metallicity using the [\ion{O}{III}]/[\ion{N}{II}] calibration of \citet{pettini2004}, and the fit parameters are
$a_1=0.27$, 
$a_2=-0.33$,
$a_3=0.71$,
and $a_4=-6.62$.
The $r_{50}$ and $12+\log(\mathrm{O/H})$ measurements were taken from the NASA-Sloan Atlas\footnote{\url{http://nsatlas.org}}.
The relation for dust temperature $T$ in Kelvin is their Equation 37
\begin{equation} \label{eq:lamperti_t}
T = b_1 \log \mathrm{SFR} + b_2 \log M_\star + b_3,
\end{equation}
where $b_1=2.91$, 
$b_2=-2.27$,
and $b_3=45.42$. 
We use these scaling relations to estimate $\beta$ and $T$ for each galaxy, and then 
estimate dust mass using Equation~\ref{eq:mdust}.

\subsection{Note regarding beam sizes}

Our measurements of CO, \hone, and dust should be interpreted as galaxy-integrated totals rather than
aperture-matched fluxes.
As noted in Section 6 of \citet{catinella2018}, although the IRAM/JCMT beams are significantly smaller than
Arecibo, it is well known that \hone\ emission extends much further than CO and dust, and so a larger beam is needed to capture all of the \hone\ emission compared to CO and dust.

\section{Analysis and Results}\label{sec:results}

\subsection{Comparing the gas and dust properties of red misfit and blue active galaxies}\label{sec:char}

\begin{figure*}
	\includegraphics[width=\linewidth]{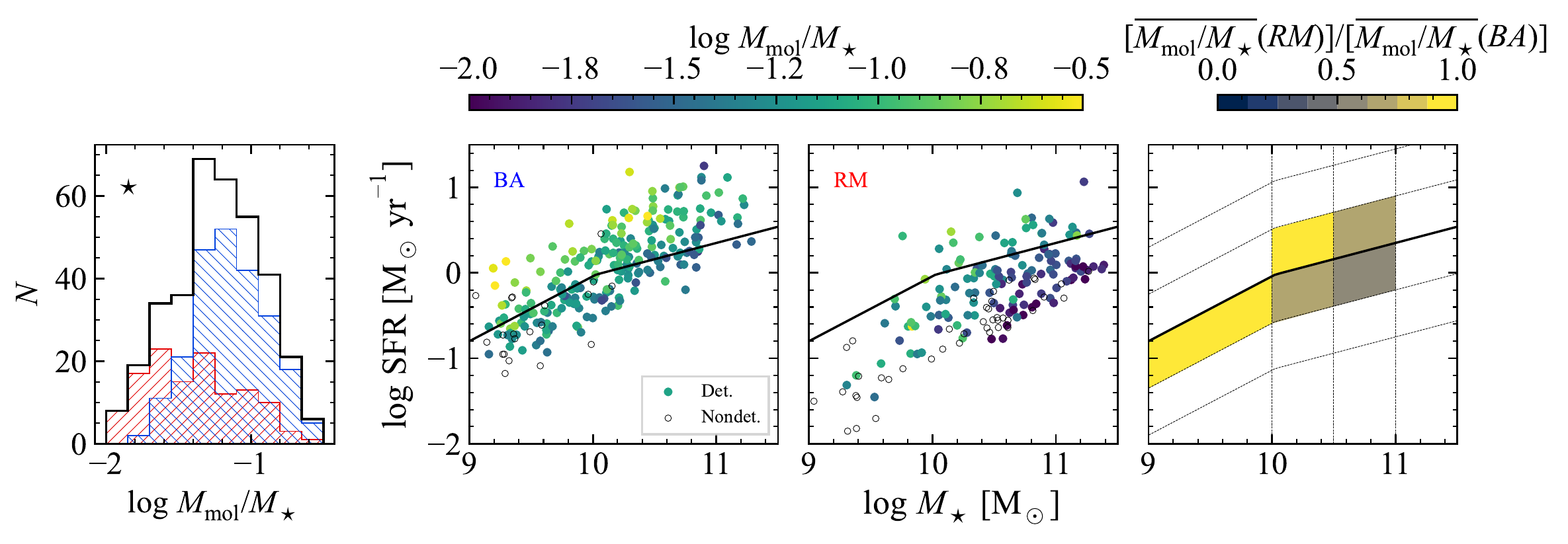}
	\includegraphics[width=\linewidth]{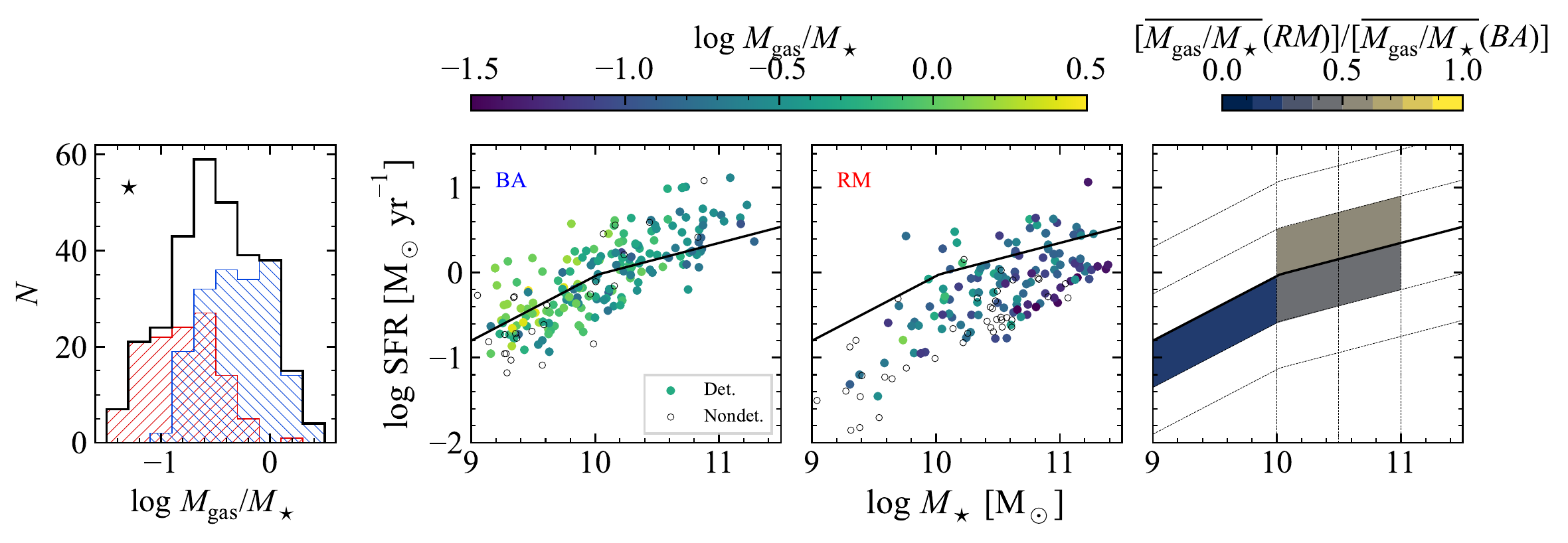}
	\caption[Gas-to-stellar mass ratios of red misfits and blue actives as histograms and in the $\log~\mathrm{SFR},\log~M_\star$ plane]{Left: Histograms of molecular (top) and total (bottom) gas mass fractions of all galaxies (black), red misfits (red histogram) and blue actives (blue histogram). Only detections are shown. The KS-test between the red misfit and blue active distributions are shown in Table~\ref{tab:ks_tests}; a ``$\star$'' symbol in the upper left of a histogram indicates that the distributions are likely different. Restricted means and the ratio of restricted means of red misfits to blue actives are shown in Table~\ref{tab:ks_tests}. Middle panels: gas mass fractions of blue actives (``BA'') and red misfits (``RM'') and their relationships to host galaxy position on the SFMS diagram. Coloured circles are detected in \htwo\ (and \hone\ where those measurements are used). Open circles were observed but not detected. The solid line is the star-forming main sequence from Table 1 of \citet{popesso2019}. The right panels show the ratio of the average gas mass fractions of red misfits to blue actives. 
    }
    \label{fig:sfms_gas}
\end{figure*}

\begin{figure*}
	\includegraphics[width=\linewidth]{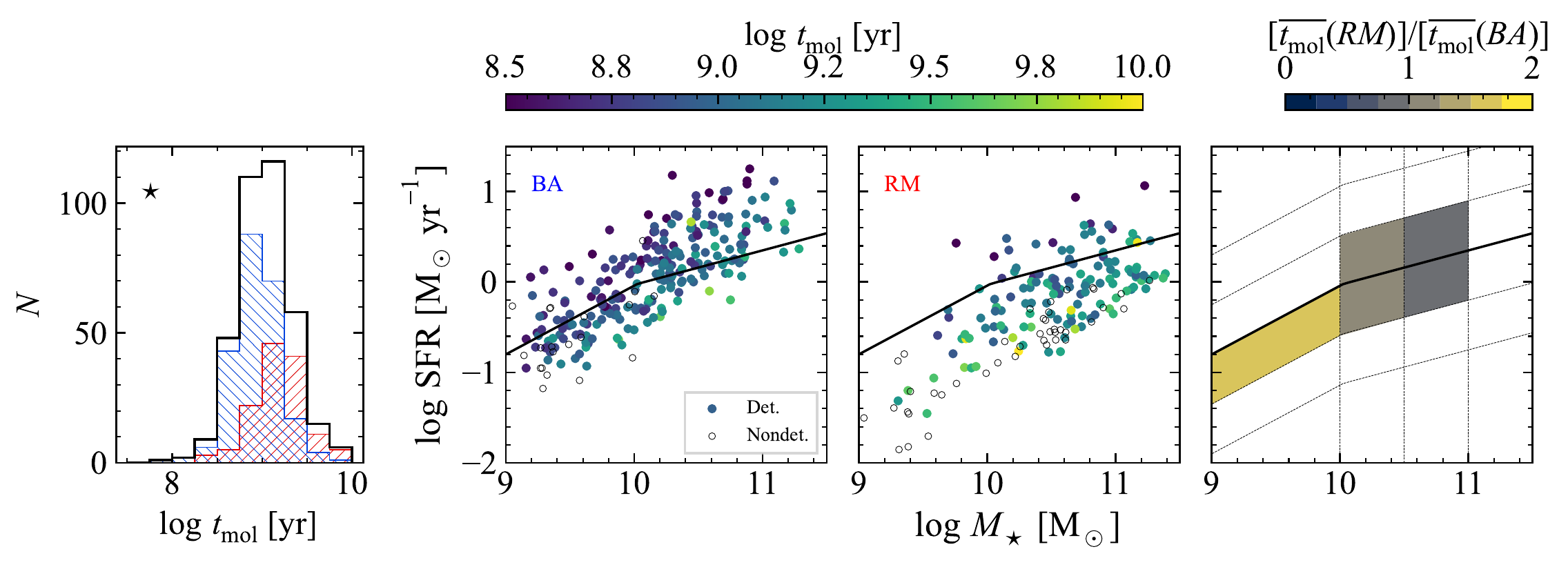}
	\includegraphics[width=\linewidth]{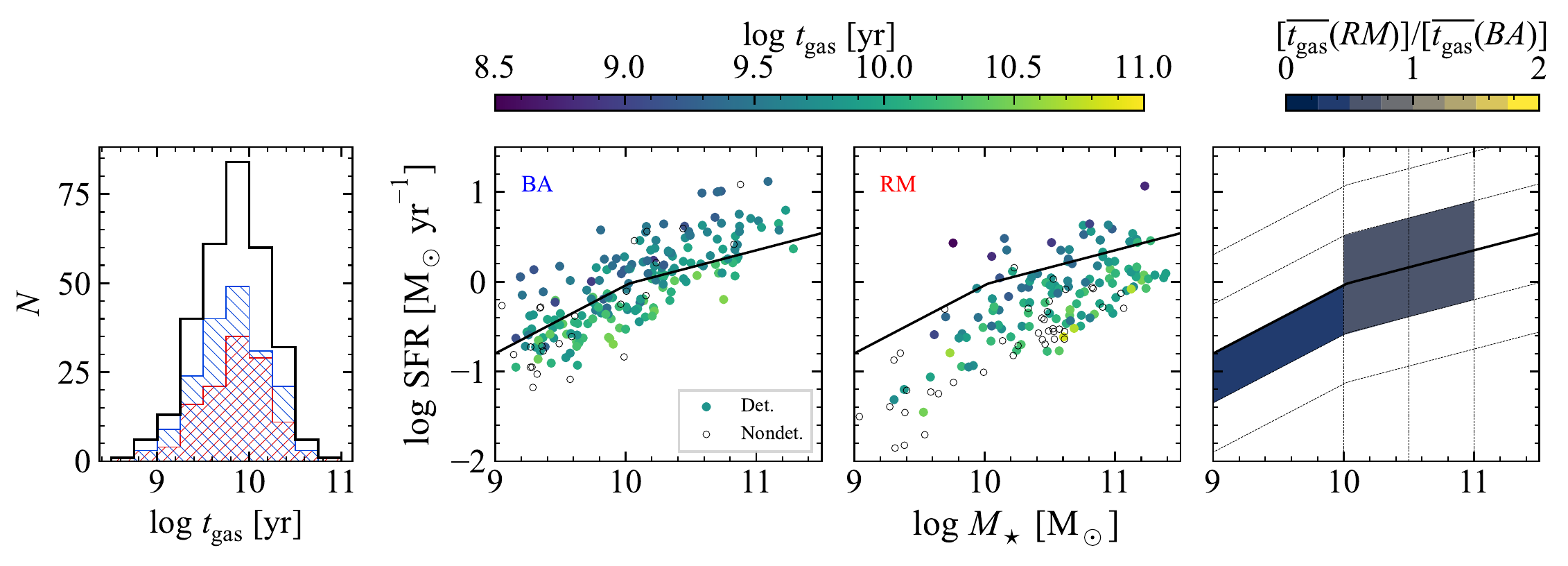}
	\caption[Gas depletion times of red misfits and blue actives as histograms and in the $\log~\mathrm{SFR},\log~M_\star$ plane]{Same as Figure~\ref{fig:sfms_gas} except with molecular gas (top) and total gas (bottom) depletion times.}
    \label{fig:sfms_tdepl}
\end{figure*}

\begin{figure*}
     \includegraphics[width=\linewidth]{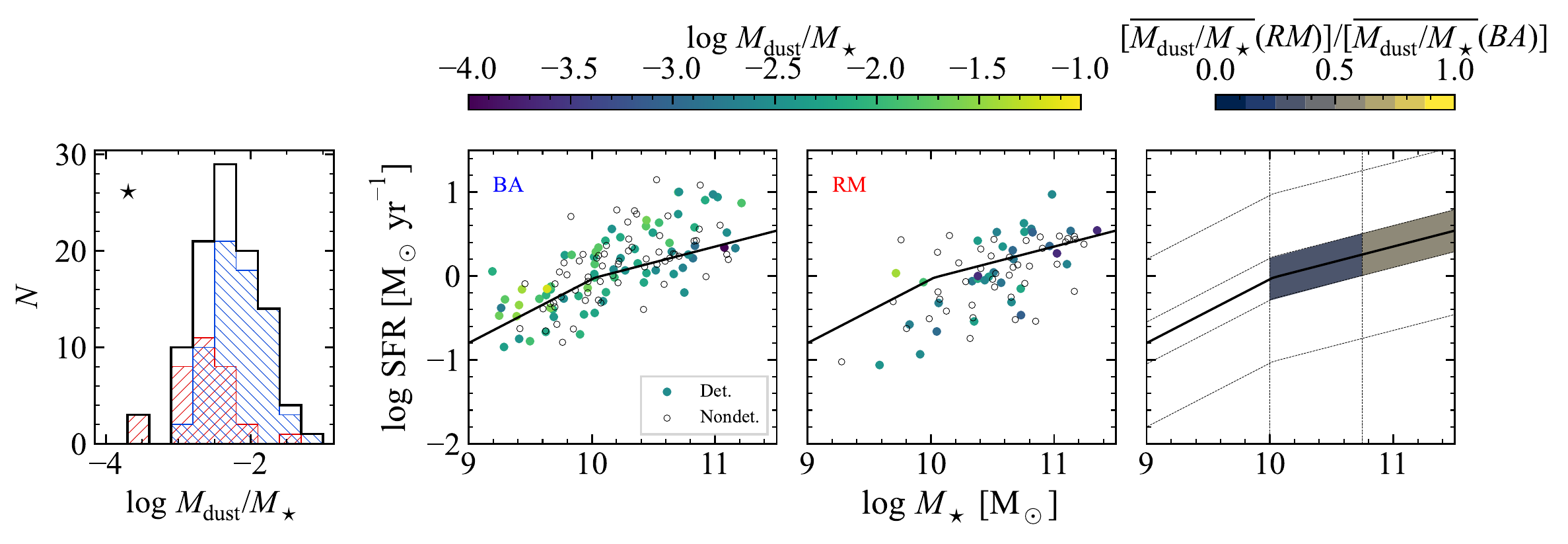}
   \includegraphics[width=\linewidth]{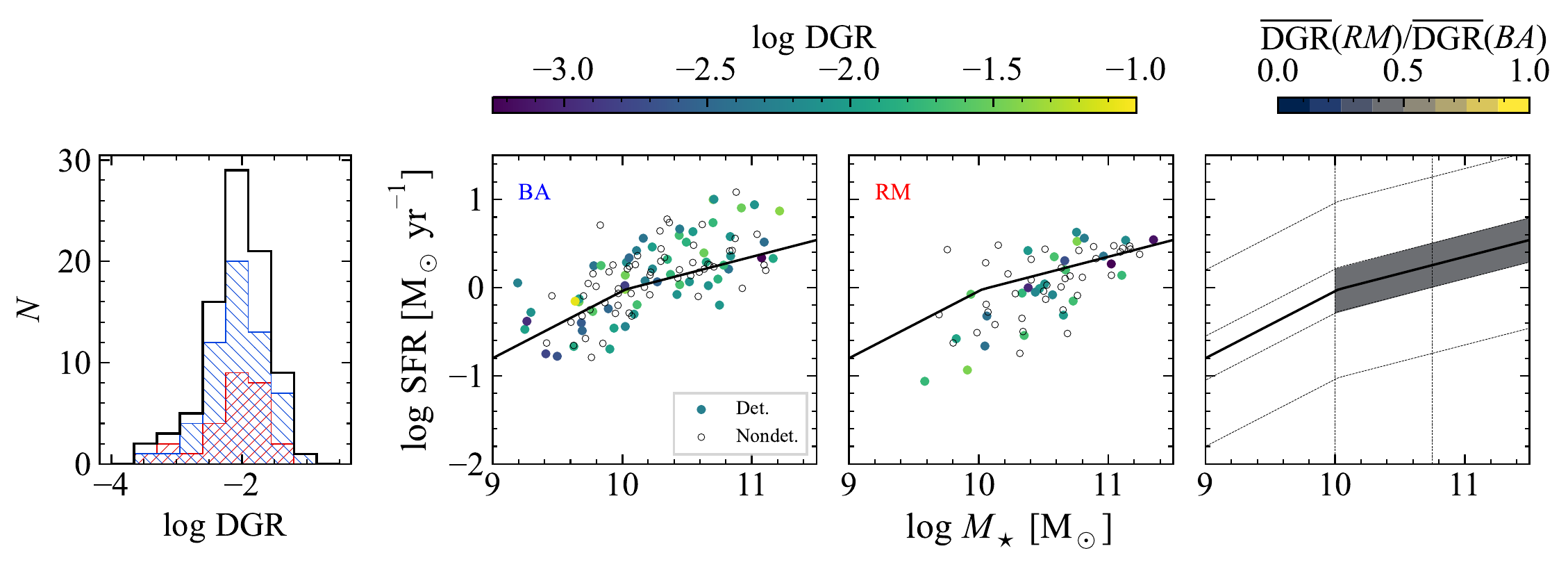}
	\caption[Dust-to-stellar mass ratios and DGRs of red misfits and blue actives as histograms and in the $\log~\mathrm{SFR},\log~M_\star$ plane]{Same as Figure~\ref{fig:sfms_gas} except with dust-to-stellar mass shown in the top row and the dust-to-gas ratio (DGR;  Equation~\ref{eq:dgr}) in the bottom row. Note that there are fewer galaxies in this figure compared to Figures~\ref{fig:sfms_gas} and~\ref{fig:sfms_tdepl} because fewer galaxies have both gas and dust measurements.
    }
    \label{fig:sfms_dust}
\end{figure*}

To better understand the nature of red misfits and their role in galaxy evolution, it is critical to
understand their gas and dust properties.
In Figures~\ref{fig:sfms_gas},~\ref{fig:sfms_tdepl}, and~\ref{fig:sfms_dust}, we show their gas masses, gas depletion times, and dust mass fractions.
We show each quantity from two perspectives in order to compare between red misfits and blue active galaxies. 
The first perspective is a comparison of distributions of detected measurements, shown in the left panels, which allows us 
to compare the properties of the entire red misfit and blue active samples. We compare the two unbinned distributions using a two-sample Kolmogorov-Smirnov (KS) test implemented in \texttt{scipy.stats.ks\_2samp}; if the resulting KS statistic is small or the p-value is large, then 
the distributions are consistent with each other.
A ``$\star$'' in the upper left of these histograms indicates that the distributions are statistically different. The results of each KS test are shown in Table~\ref{tab:ks_tests}. 

The second perspective shows gas mass properties, depletion times, and dust mass fractions in the SFR-$M_\star$ plane 
(right panels of Figures~\ref{fig:sfms_gas}, \ref{fig:sfms_tdepl}, and \ref{fig:sfms_dust}). 
With this method we can explore how dust and gas properties for red misfits and blue actives depend on 
their position relative to the SFMS.
For example, in the right panel of the first row of Figure~\ref{fig:sfms_gas}, the colour of each bin shows the average $M_\mathrm{mol}/M_\star$ for red misfits divided by 
that of blue actives in that bin.
Viewing the sample this way allows us to examine the differences
in gas and dust properties while controlling for the fact that red misfits and blue actives are distributed differently in the $\log \mathrm{SFR}$-$\log M_\star$ plane.

In Table~\ref{tab:ks_tests}, we also show the restricted mean and standard error of each quantity for red misfits and blue actives separately, taking non-detections into account.
This was done using the Kaplan-Meier estimator (implemented in the \texttt{lifelines} Python package),
from which we extract a restricted mean and standard error. 
The Kaplan-Meier estimator is a survival analysis algorithm which 
estimates the probability distribution of a quantity when measurements of this quantity contain both detections and non-detections. The ``restricted mean'' is an estimate of the mean of the true distribution. The restricted mean is defined as the integral of the estimated survival function up to the largest detected data point. A recent application of the Kaplan-Meier estimator to molecular gas measurements of galaxies can be found in \citet{mok2016}.

We compare the relative amount of cold gas in these two populations 
through two quantities: the molecular-to-stellar mass ratio $M_\mathrm{mol}/M_\star$, and the total gas to stellar mass ratio $M_\mathrm{gas}/M_\star$,
shown in Figure~\ref{fig:sfms_gas}. The distributions on the left and the KS test results (Table~\ref{tab:ks_tests})
indicate that red misfits
tend to have lower gas mass fractions than blue active galaxies. 
The middle two panels show that red misfits and blue actives are distributed differently in the $\log \mathrm{SFR}$-$\log M_\star$ plane (red misfits tend to lie below the SFMS especially at high stellar masses), and that
the gas fractions vary within this space. To compare the average properties as functions of position in the $\log \mathrm{SFR}$-$\log M_\star$ plane,
we computed the average gas fractions (detections only) in two-dimensional bins of $\log \mathrm{SFR}$ and $\log M_\star$ (right column). We require a minimum of three red misfits and three blue actives per bin. 
In Figure~\ref{fig:sfms_gas}, aside from the lowest-$\log M_\star$ bin and the bin between $10\leq \log M_\star\leq 10.5$ which lies above the SFMS, red misfits have lower molecular gas and total gas mass fractions than blue actives. In the two exceptional bins, red misfits have higher molecular gas mass fractions, and lower total gas mass fractions than blue actives. These two exceptional bins contain few red misfits, and so may not adequately represent the whole population.
The restricted mean gas fractions for red misfits 
and blue actives are shown in Table~\ref{tab:ks_tests}. The ratio of red misfit to blue active gas fractions are significantly less than unity, which supports our findings from detections alone (left panel of Figure~\ref{fig:sfms_gas}). This indicates that red misfits have lower total gas content and molecular gas content relative to blue actives.

Next we compare the molecular and total gas depletion times (Figure~\ref{fig:sfms_tdepl}). 
Based on the KS test comparing the red misfit and blue active distributions (Table~\ref{tab:ks_tests}),  the $t_\mathrm{mol}$ 
distributions are significantly different, while the $t_\mathrm{gas}$ distributions are not. 
This result is also supported by the ratios of the restricted means -- compared to blue actives, the mean $t_\mathrm{mol}$ of red misfits is slightly larger  and the difference is statistically significant.
The mean $t_\mathrm{gas}$ of both populations are not significantly different.
This indicates that the molecular gas will be depleted more slowly in red misfits than blue actives, but the total
gas reservoirs deplete at nearly the same rates. 
From the other panels of Figure~\ref{fig:sfms_tdepl} there are no
clear trends in the ratios of $t_\mathrm{mol}$ or $t_\mathrm{gas}$.

Finally we show the dust-to-stellar mass ratios $M_\mathrm{dust}/M_\star$ and the dust-to-gas ratios
\begin{equation}\label{eq:dgr}
\mathrm{DGR} \equiv \frac{M_\mathrm{dust}}{M_\mathrm{gas}},
\end{equation}
in Figure~\ref{fig:sfms_dust}. The $M_\mathrm{dust}/M_\star$ distributions (top row of Figure~\ref{fig:sfms_dust})
are different based on a KS test (Table~\ref{tab:ks_tests}), with the blue actives having significantly higher values, while the DGR distributions do not show a significant difference. This finding is also supported by the fact that 
red misfits have significantly lower restricted mean $M_\mathrm{dust}/M_\star$ value than blue actives. Red misfits have a smaller restricted mean DGR than blue actives, but this is not statistically significant (only $2\sigma$).
In the SFMS plane (middle panels of Figure~\ref{fig:sfms_dust}), aside from a few bins with small number of galaxies in them, 
red misfits tend to have lower $M_\mathrm{dust}/M_\star$ in all areas of this plane.
Dust-to-gas ratios also do not show any strong differences between red misfits and blue actives in this plane. 
These results indicate that red misfits contain \textit{less} dust than 
blue active galaxies, rather than more dust as one might initially expect based on their red optical colours.

\begin{table*}
\caption[Statistical comparisons of gas- and dust-based quantities between red misfits and blue active galaxies]{Statistical comparisons of gas- and dust-based quantities between red misfits and blue active galaxies.}
\label{tab:ks_tests}
\begin{center}
\begin{tabular}{lllcllcc}
\hline
Quantity & KS statistic & p-value & Different? & \multicolumn{2}{c}{Restricted mean}  & log(RM/BA) & Figure \\
 \hline
(1) & (2) & (3) & (4) &  \multicolumn{2}{c}{(5)} & (6) & (7) \\
 \hline
 & & & & \multicolumn{1}{c}{RM} & \multicolumn{1}{c}{BA} &  & \\
\hline
$\log\> M_\mathrm{mol}/M_\star$ & 0.428 & $1.40\times 10^{-14}$ & Y & $-1.61\pm0.03$ & $-1.16\pm0.02$ & $-0.46\pm0.04$ & \ref{fig:sfms_gas} \\
$\log\> M_\mathrm{gas}/M_\star$ & 0.592 & $3.33\times 10^{-16}$ & Y & $-0.94\pm0.03$ & $-0.33\pm0.02$ & $-0.61\pm0.04$ & \ref{fig:sfms_gas} \\
$\log\> t_\mathrm{mol}$ [yr] & 0.409 & $2.70\times 10^{-13}$ & Y & $9.10\pm0.03$ & $8.91\pm0.02$ & $0.20\pm0.03$ & \ref{fig:sfms_tdepl} \\
$\log\> t_\mathrm{gas}$ [yr] & 0.116 & $2.47\times 10^{-1}$ & N & $9.76\pm0.03$ & $9.77\pm0.03$ & $-0.01\pm0.05$ & \ref{fig:sfms_tdepl} \\
$\log\> M_\mathrm{dust}/M_\star$ & 0.536 & $1.79\times 10^{-6}$ & Y & $-2.87\pm0.07$ & $-2.20\pm0.05$ & $-0.67\pm0.09$ & \ref{fig:sfms_dust} \\
$\log\> \mathrm{DGR}$ & 0.112 & $9.44\times 10^{-1}$ & N & $-2.32\pm0.09$ & $-2.07\pm0.05$ & $-0.2\pm0.1$ & \ref{fig:sfms_dust} \\
\hline
\multicolumn{8}{l}{(2) Kolmogorov-Smirnov (KS) statistic comparing red misfits and blue actives. This includes detections only, by definition (\S~\ref{sec:char}).}\\
\multicolumn{8}{l}{(3) p-value corresponding to the KS statistic.}\\
\multicolumn{8}{l}{(4) Are the distributions statistically different based on the KS statistic (Y/N)? Y if KS $\geq 0.4$ and $p \ll 1$. }\\
\multicolumn{8}{l}{(5) Kaplan-Meier restricted mean and standard error (\S~\ref{sec:char}).}\\
\multicolumn{8}{l}{(6) Ratio of the restricted mean of red misfits to blue actives, in logarithmic units.}\\
\end{tabular}
\end{center}
\end{table*}

\subsection{Scaling relations}\label{sec:scaling_relations}

A key question in understanding the evolution of star-forming galaxies is what
drives the scatter about the SFMS.
Recently, assessing the relative  
importance of gas depletion time and gas mass fraction in driving the scatter about the SFMS has been a major focus \citep[see, e.g.,][]{saintonge2016, lin2019a, ellison2020, feldmann2020, sanchez2021}. Here we explore whether there are differences in how depletion times and gas 
mass fractions of red misfits and blue actives correlate with offset from the SFMS.
To answer these questions, we plot $\log~t_\mathrm{mol}$, $\log~t_\mathrm{gas}$,  $\log~M_\mathrm{mol}/M_\star$, and  $\log~M_\mathrm{gas}/M_\star$ versus offset 
from the SFMS.
The offset from the star forming main sequence is defined as
\begin{equation}\label{eq:delta_ms}
\Delta \mathrm{MS} = \log \mathrm{SFR}(M_\star) - \log \mathrm{SFR_{MS}}(M_\star),
\end{equation}
where $\log \mathrm{SFR}(M_\star)$ is the SFR of a galaxy with stellar mass $M_\star$ and
$\log \mathrm{SFR_{MS}}(M_\star)$ is the star forming main sequence \citep{popesso2019} at the same stellar mass
\begin{equation}\label{eq:sfms}
\log \mathrm{SFR_{MS}}(M_\star) = 0.38 \log M_\star - 3.83.
\end{equation}
We adopt this particular definition of the SFMS because it was derived from the same 
SFR and $M_\star$ measurements that we use here, namely those from the GSWLC-M2 catalog.

\begin{figure*}
	\includegraphics[width=0.37\linewidth]{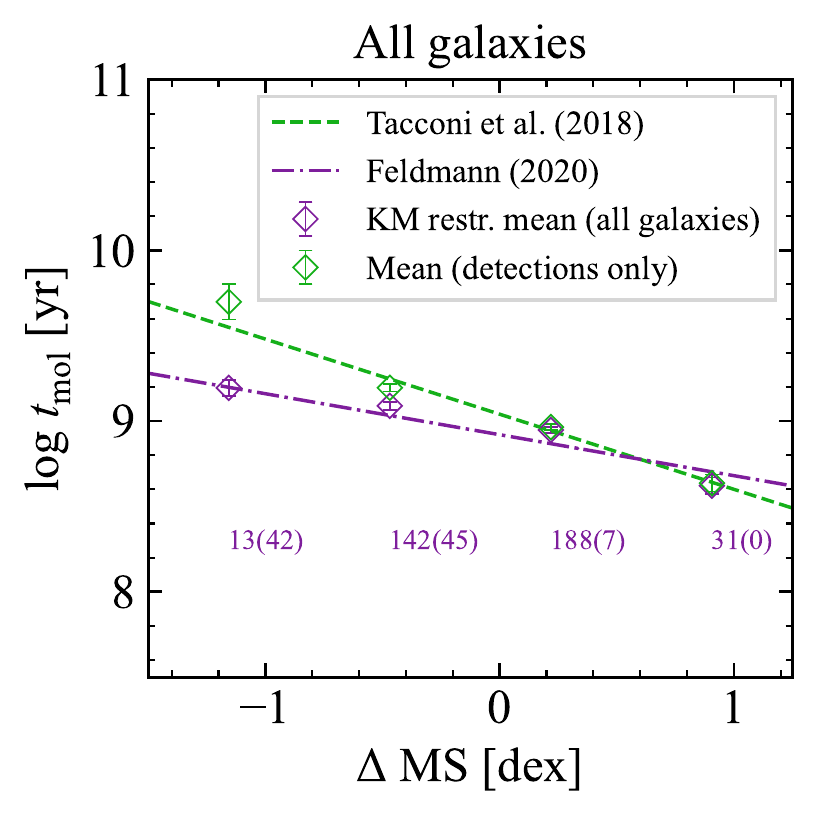}\includegraphics[width=0.37\linewidth]{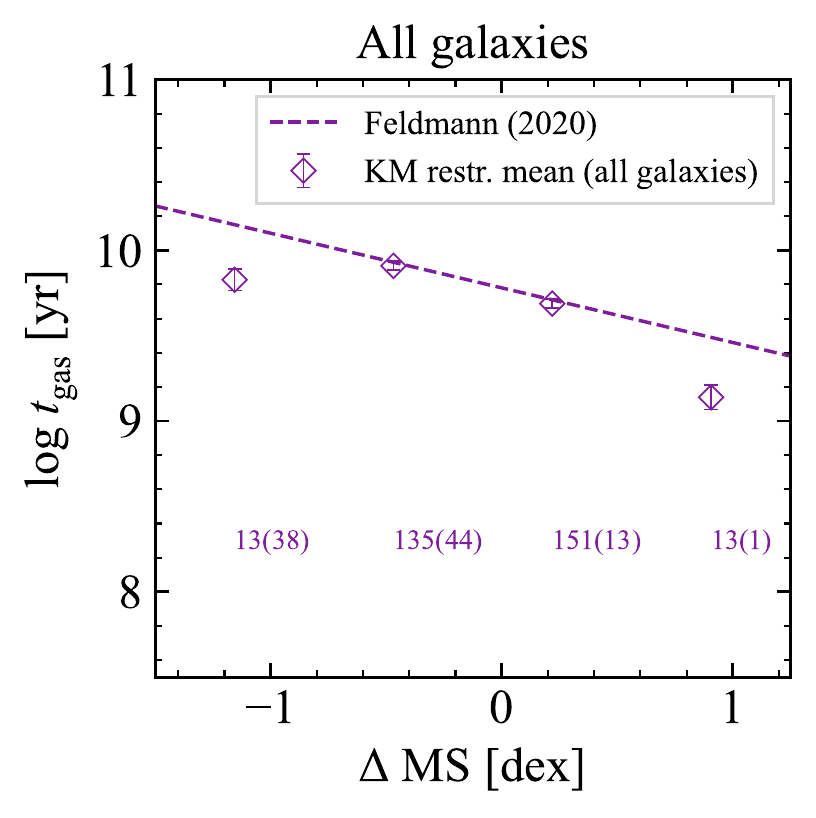}
	\includegraphics[width=0.37\linewidth]{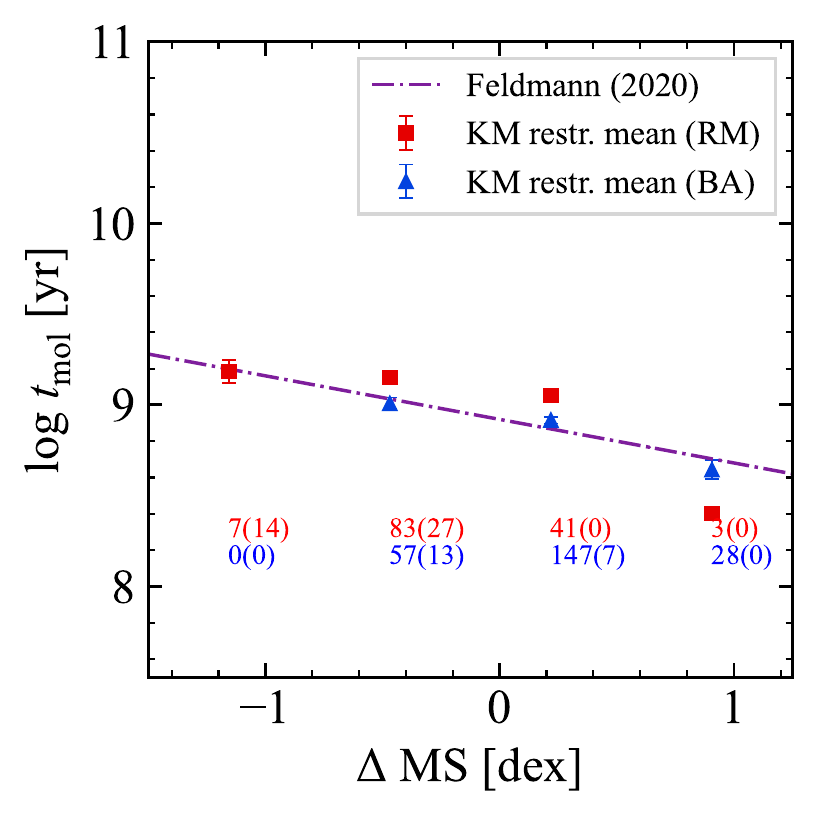}\includegraphics[width=0.37\linewidth]{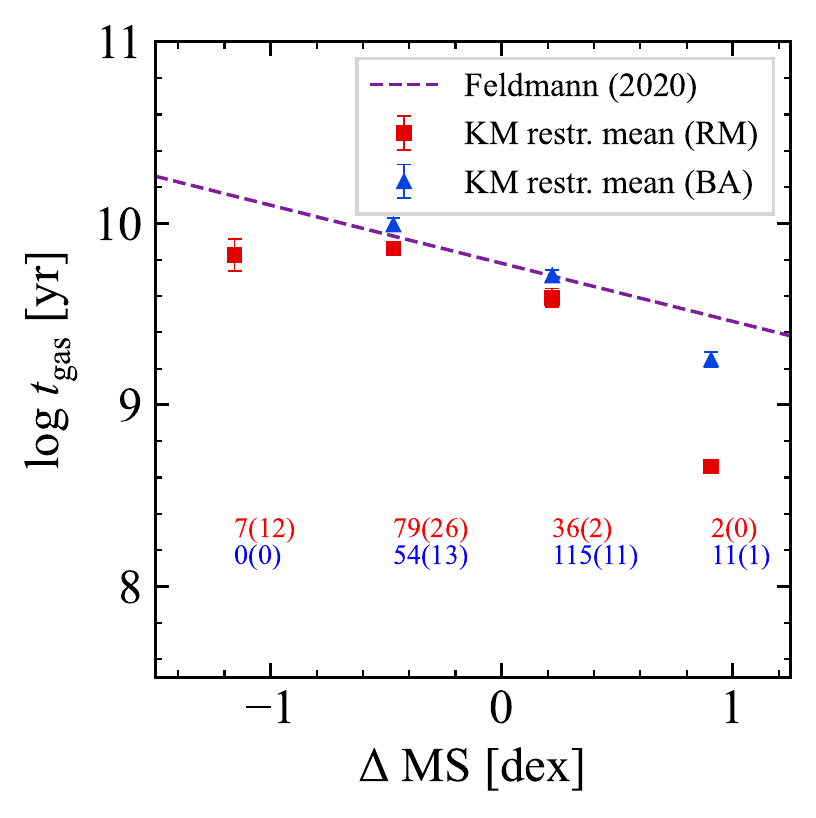}
\includegraphics[width=0.37\linewidth]{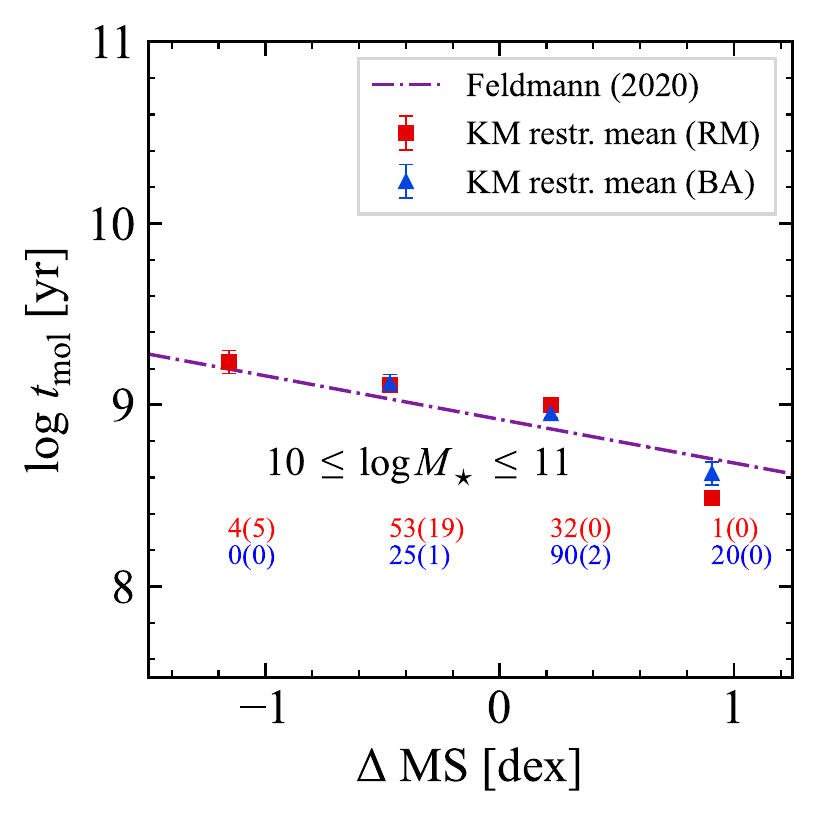}\includegraphics[width=0.37\linewidth]{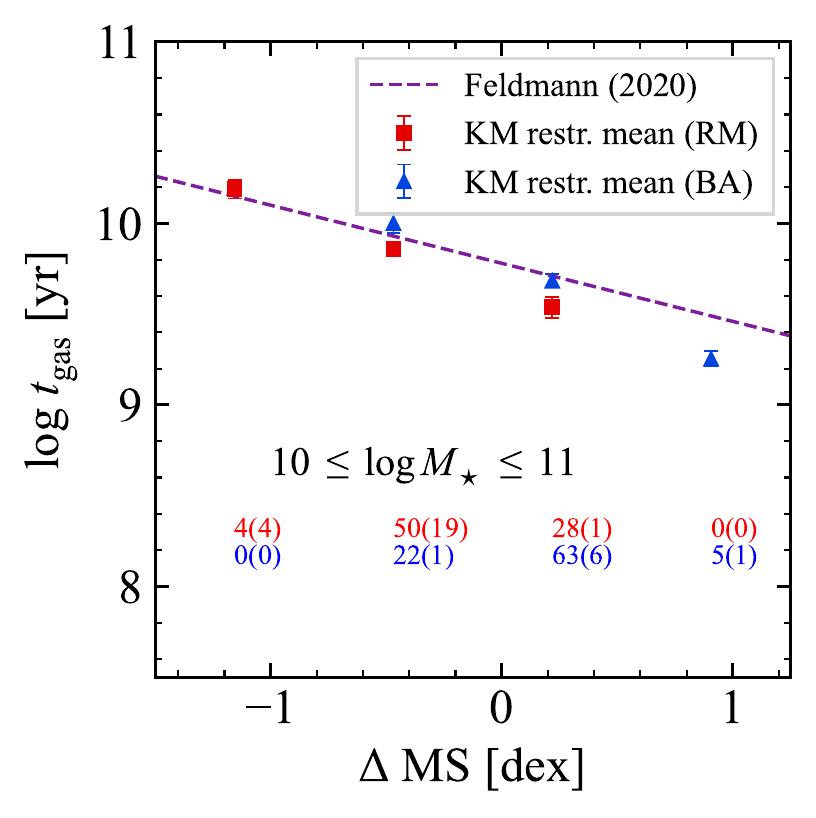}
	\caption[]{Molecular (left) and total (right) gas depletion times versus offset from the star forming main sequence (Equation~\ref{eq:delta_ms}). The purple dot-dashed lines are from \citet{feldmann2020}, which are fits to molecular (left) and total (right) gas depletion times from xCOLD GASS (+xGASS) after taking non-detections into account. The green dashed line is the fit to $z=0-4$ galaxies from the PHIBSS survey \citep{tacconi2018} for comparison. The Kaplan-Meier median is the value where the cumulative distribution reaches 0.5. The top row shows all galaxies in our sample. One can see that our results agree well with \citet{tacconi2018} when we ignore non-detections and with \citet{feldmann2020} when we include non-detections. The middle row shows red misfits and blue actives separately. 
	The number of detections are shown below each data point, with the number of non-detections shown in brackets.
	One can see that, when there is enough data to compute a median ($\geq 5$ points), the molecular and total gas depletion times of red misfits and blue actives are similar, although slightly (but statistically significantly) larger for red misfits. The bottom row is the same as the middle row except only showing galaxies with stellar masses between $10^{10}$ and $10^{11}~M_\odot$, where both depletion times become more similar between red misfits and blue actives.
    }
    \label{fig:tdepl_delta_ms}
\end{figure*}

In Figure~\ref{fig:tdepl_delta_ms} we show molecular gas $t_\mathrm{mol}$ (left column) and total gas $t_\mathrm{gas}$ depletion times (right column) versus $\Delta \mathrm{MS}$. 
The restricted mean and standard error (see Section~\ref{sec:char}) of each quantity is computed in bins of $\Delta \mathrm{MS}$.
As a test of our method, in the top row we compare our relations with those from \citet{feldmann2020}, which shows good agreement. There are some notable differences between their study and ours: in \citet{feldmann2020} the xCOLD GASS sample was used, whereas here we are using a larger sample and a slightly different definition of the SFMS; they took non-detections into account using a method that is different than ours \citep[LeoPy;][]{feldmann2019}. 
We also compare our $\log~t_\mathrm{mol}$-$\Delta~\mathrm{MS}$ relationship using the average 
of detections only with the relationship found by \citet{tacconi2018} with the IRAM Plateau de Bure high-$z$ blue sequence CO(3-2) survey \citep[PHIBSS;][]{tacconi2013}, who did not incorporate non-detections. Although their sample is notably different than ours in terms of redshift ($z\sim 1$ to 2 versus $z\sim 0$ in our work), this plot shows that our results are in good agreement with theirs. Overall, these relationships show that as galaxies 
move from above to below the main sequence their gas is used up more slowly (e.g. $t_\mathrm{gas}$ increases).

Having confirmed that our results for the sample as a whole agree with previous work, we
move on to study these trends for red misfits and blue actives separately in the middle and bottom rows of Figure~\ref{fig:tdepl_delta_ms}. 
In the middle row, we see that both populations follow similar 
trends to the population as a whole; however red misfits have higher
 $t_\mathrm{mol}$ values than blue actives below the SFMS and up to  
 $\Delta~\mathrm{MS}\sim 0.2$ dex. This is in line with our previous result 
in Table~\ref{tab:ks_tests}, which showed that red misfits have longer 
$t_\mathrm{mol}$ and similar $t_\mathrm{gas}$ compared to blue actives. Here, however, we see that this difference is primarily coming from galaxies on and below the main sequence. In the bottom row we show the same as the middle row except only galaxies with
$10\leq \log~M_\star\leq 11$, which is where both red misfits and blue actives are well-sampled.  
The red and blue points are closer together than in the middle row.
This result indicates that the differences seen in the left panel of the second row 
are largely coming from galaxies outside of this stellar mass range.

\begin{figure*}
	\includegraphics[width=0.37\linewidth]{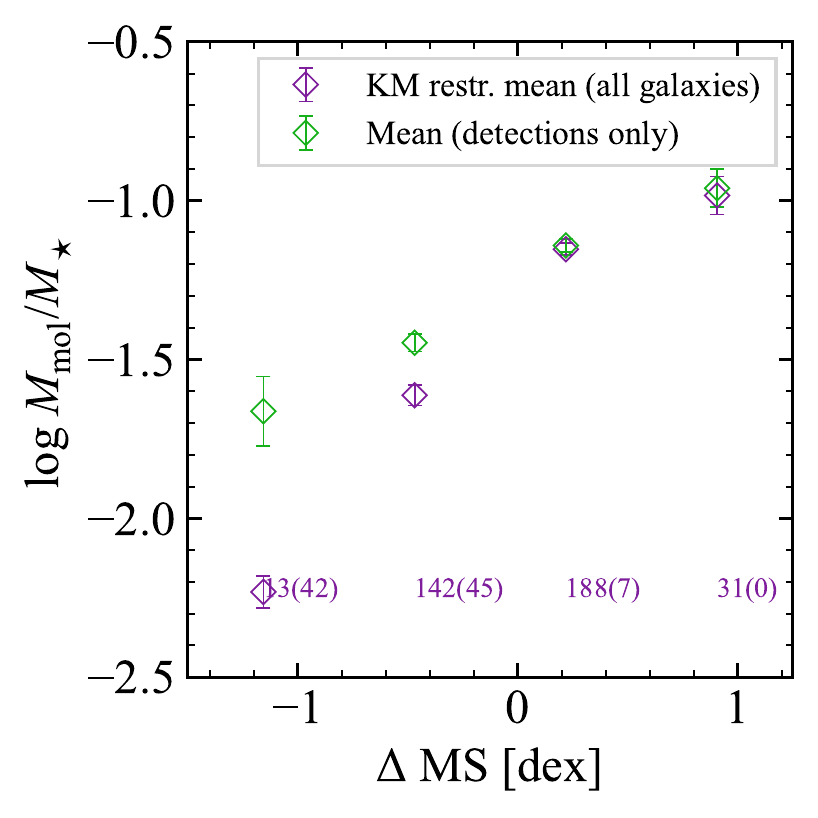}\includegraphics[width=0.37\linewidth]{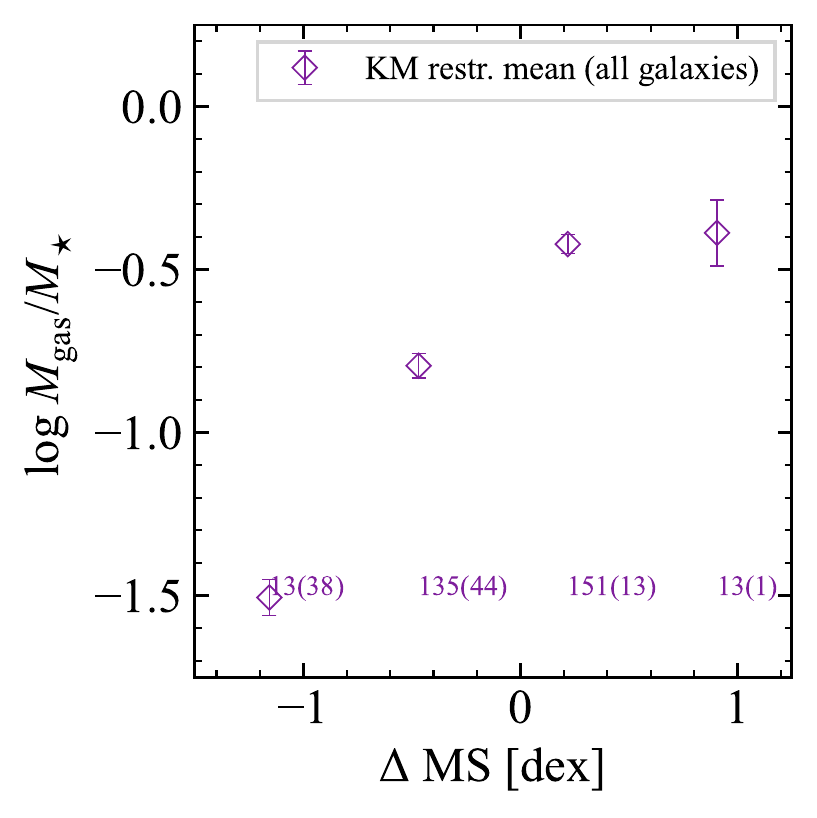}
	\includegraphics[width=0.37\linewidth]{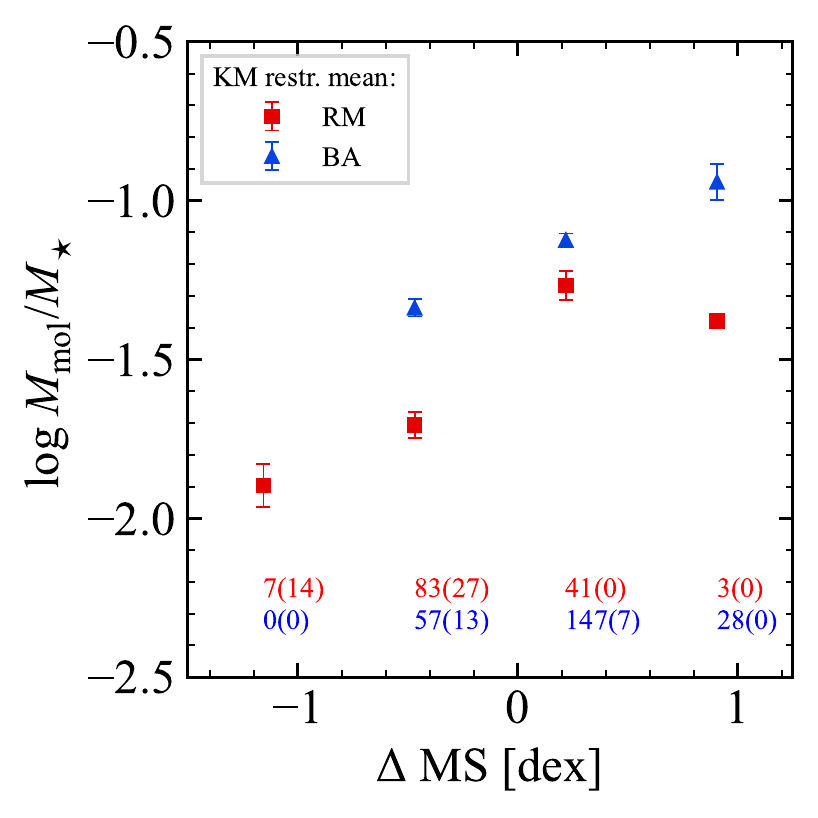}\includegraphics[width=0.37\linewidth]{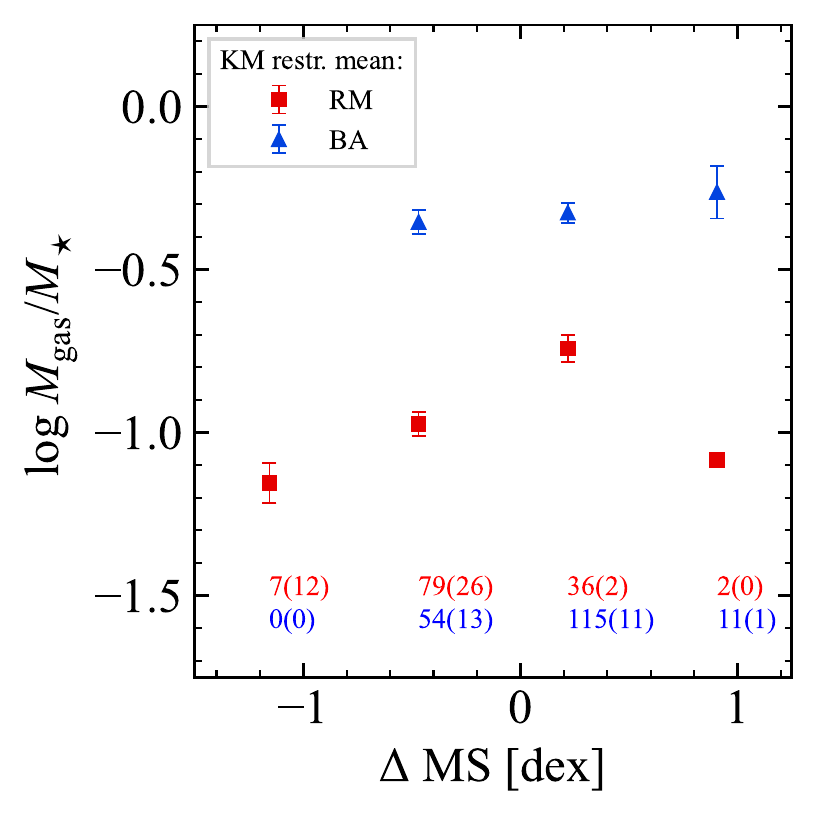}
	\includegraphics[width=0.37\linewidth]{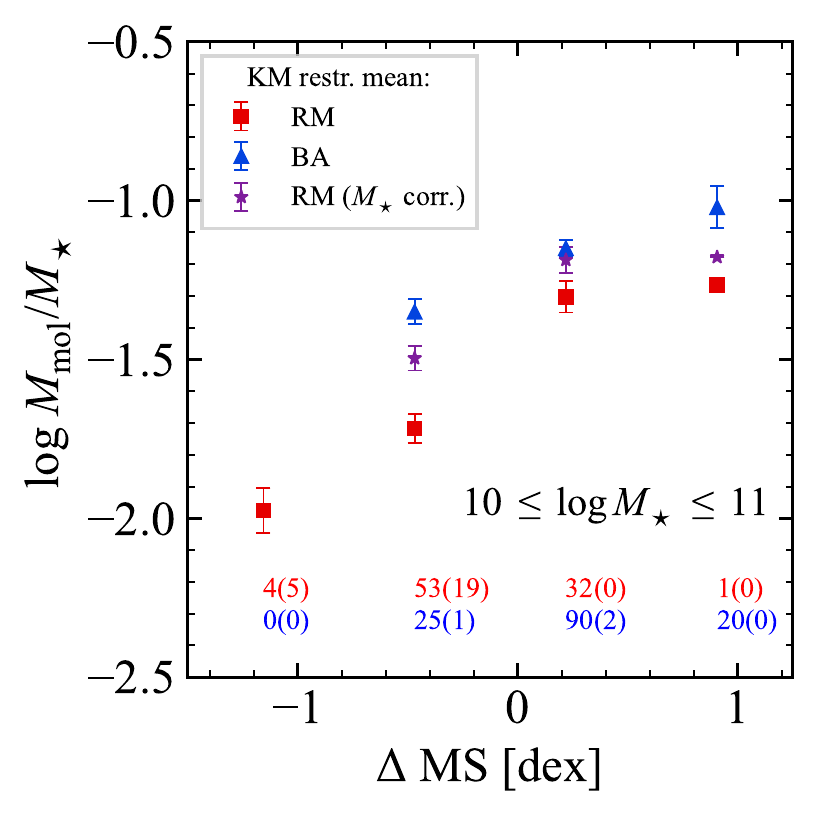}\includegraphics[width=0.37\linewidth]{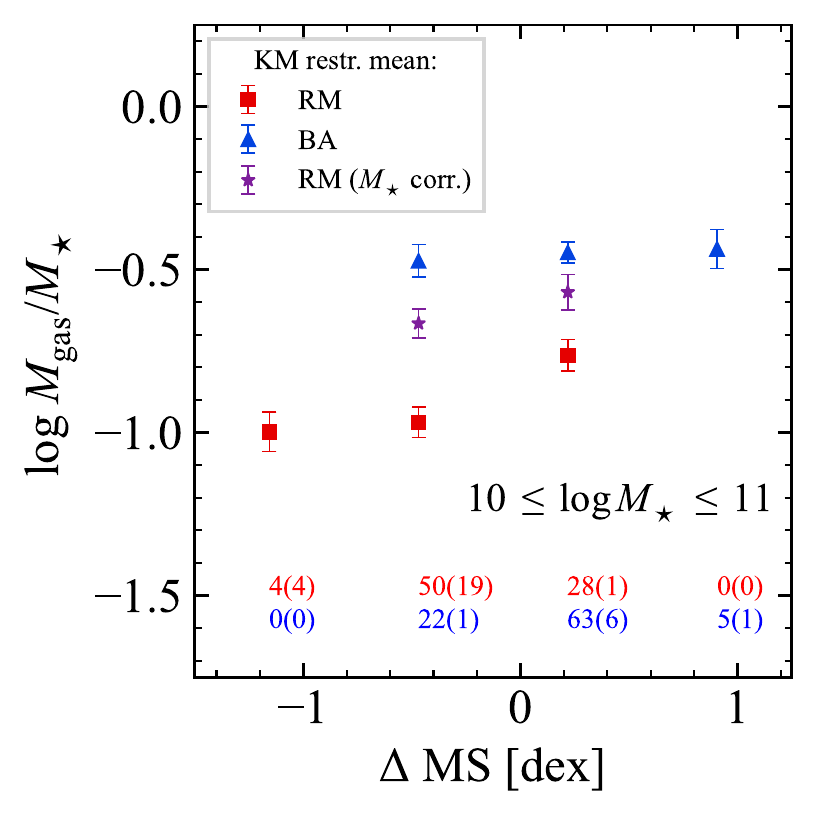}
	\caption[Molecular and total gas mass fractions times versus offset from the star forming main sequence]{Molecular (left) and total (right) gas mass fractions versus offset from the star forming main sequence (Equation~\ref{eq:delta_ms}). The top row shows all galaxies in our sample. One can see that across the main sequence, molecular gas mass fraction rises, while the total gas mass fraction rises and then remains constant. The middle row shows red misfits and blue actives separately. The number of detections are shown below each data point, with the number of non-detections shown in brackets. One can see that, in contrast to Figure~\ref{fig:tdepl_delta_ms}, red misfits tend to have lower molecular and total gas fractions below and on the main sequence. The bottom row is the same as the middle row except only showing galaxies with stellar masses between $10^{10}$ and $10^{11}~M_\odot$, where both gas mass fractions change slightly but remain significantly different  between red misfits and blue actives. The ``$M_\star$ corr.'' points in the bottom row are gas fractions of red misfits, corrected for the difference in median stellar mass between red misfits and blue actives in each $\Delta$ MS bin (see Section~\ref{sec:mstar_bias}).}
    \label{fig:fgas_delta_ms}
\end{figure*}

In Figure~\ref{fig:fgas_delta_ms} we show molecular gas mass fractions $M_\mathrm{mol}/M_\star$ and total gas $M_\mathrm{gas}/M_\star$ mass fractions versus $\Delta \mathrm{MS}$. We use the same survival analysis approach as above to take non-detections into 
account in each $\Delta \mathrm{MS}$ bin. 
Note that we do not have curves from the literature to show for comparison. 
The top row shows that $M_\mathrm{mol}/M_\star$ increases as galaxies move from 
below to above the main sequence, 
while  $M_\mathrm{gas}/M_\star$ increases with increasing $\Delta \mathrm{MS}$ up to $\Delta \mathrm{MS} \sim -0.5$ dex 
and then remains constant at $\sim 0.5$ dex.
In the middle row, we see that the trends for red misfits and blue actives are different, especially below the main sequence. 
Red misfits have significantly lower $M_\mathrm{mol}/M_\star$ and  $M_\mathrm{gas}/M_\star$ relative to blue active galaxies, although the differences become less significant on and above the main sequence. 
This indicates that red misfits are quite gas-poor, despite their relatively similar gas depletion times compared to blue actives (Figure~\ref{fig:sfms_tdepl}). 
This result is echoed by the comparison of restricted means of these properties for red misfits and blue actives altogether (Table~\ref{tab:ks_tests}): red misfits have significantly lower gas mass fractions than blue actives.
In the bottom row of Figure~\ref{fig:fgas_delta_ms}, we show the same as the middle row but only for galaxies with
$10\leq \log~M_\star\leq 11$, which is where both red misfits and blue actives are well-sampled. Relative differences between red misfits and blue actives decrease slightly, indicating that controlling for stellar mass reduces 
differences between the populations. We explore this further in Section~\ref{sec:mstar_bias}.

\subsection{The $\log~\mathrm{SFR}$-$\log~M_\mathrm{mol}$ relationship, and the molecular gas main sequence}\label{sec:ks_mgms}

\begin{figure*}
\centering
	\includegraphics[width=\linewidth]{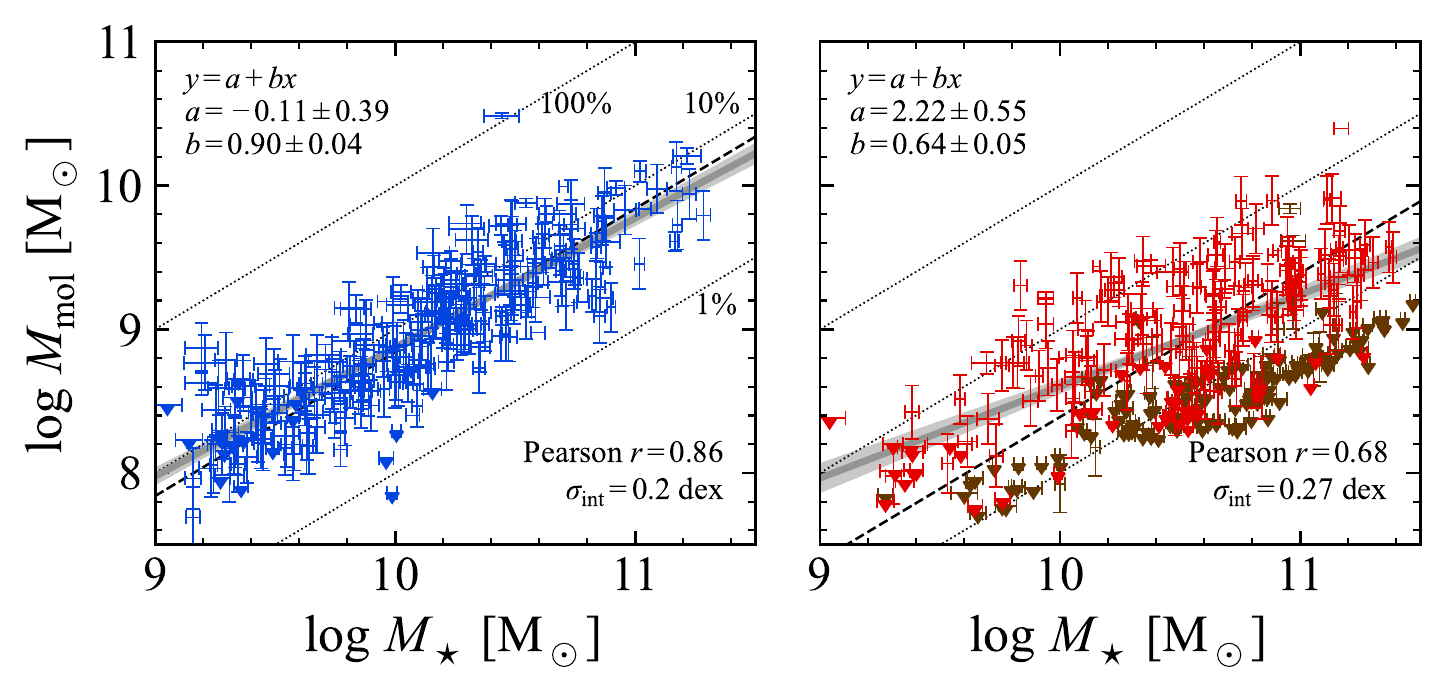}
	\includegraphics[width=\linewidth]{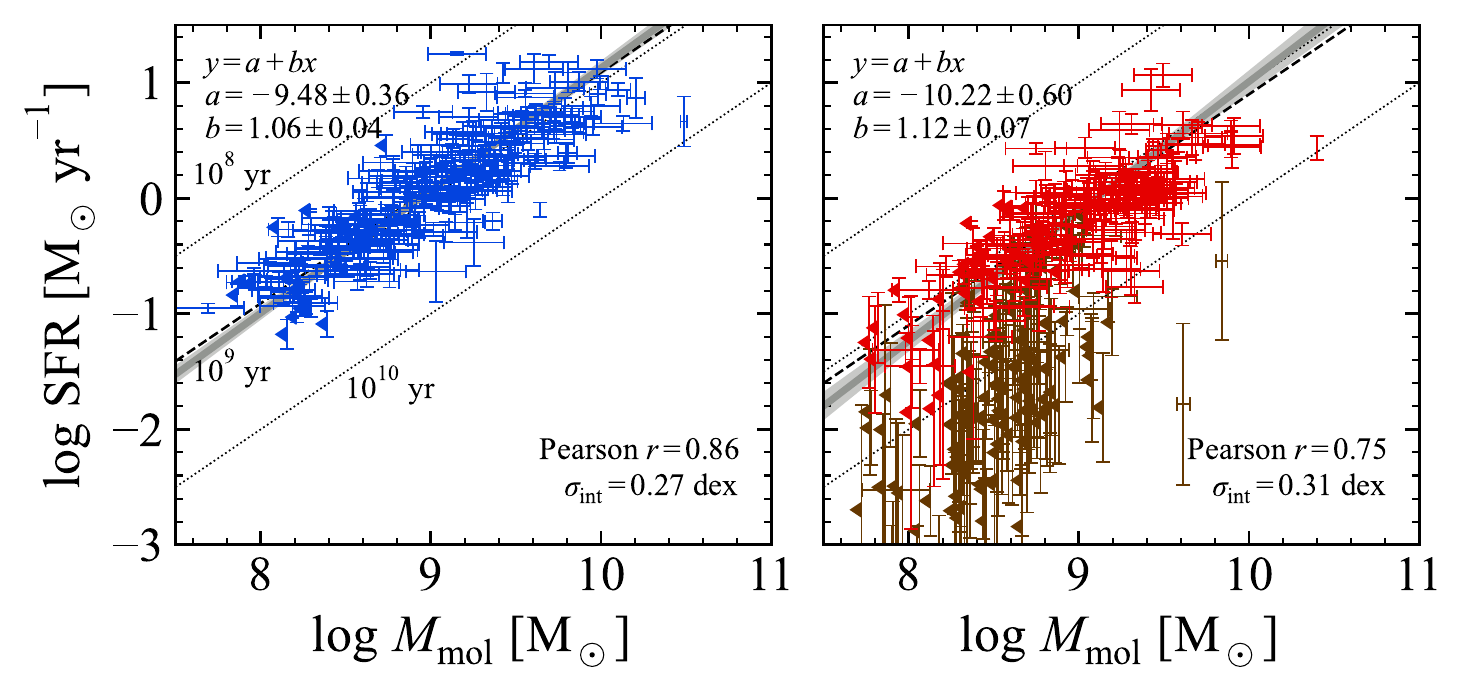}
	\caption[The MGMS and $\log~\mathrm{SFR}$-$\log~M_\mathrm{mol}$  relation of red misfits and blue actives]{The molecular gas main sequence (top) and $\log~\mathrm{SFR}$-$\log~M_\mathrm{mol}$ relationship with molecular gas (bottom) for blue active (left), red misfit galaxies (red points on the right), and red-and dead galaxies (brown points on the right). The linear fits in all panels were performed using \texttt{linmix} \citep{kelly2007} which incorporates uncertainties in $x$ and $y$, and upper limits in $y$. 
In each panel the intrinsic scatter $\sigma_\mathrm{int}$ \citep[see Appendix C of][]{chown2021} is indicated.
In the bottom row the fits were done with $\log \mathrm{SFR}$ on the x-axis, and the best-fit equations were inverted to match how this relationship is usually shown with gas on the x-axis and SFR on the y-axis. In the top row, the dotted lines represent constant $M_\mathrm{mol}/M_\star$ (1, 10, and 100 per cent), and the dashed lines correspond to the restricted mean $M_\mathrm{mol}/M_\star$ for blue actives and red misfits from Table~\ref{tab:ks_tests}. The bottom rows show lines of constant $t_\mathrm{mol}$, and the restricted means from  Table~\ref{tab:ks_tests}. These plots show that red misfits tend to have lower $M_\mathrm{H_2}/M_\star$ fractions, and slightly longer $t_\mathrm{mol}$. The slope of the MGMS is significantly flatter for red misfits than for blue actives. Red and dead galaxies (brown points) are shown for comparison in the right panels. Although most of their gas masses are upper limits, it is apparent that red and dead galaxies have longer gas depletion times and lower gas mass fractions than red misfits.
    }
    \label{fig:ks_mgms}
\end{figure*}

Stellar mass, SFR, and molecular gas are correlated with each other, as shown by the SFMS and the Kennicutt-Schmidt relation \citep[SFR surface density vs. cold gas surface density;][]{kennicutt1989, kennicutt2007, bigiel2008, leroy2008, leroy2013}.
Recent work has introduced the ``molecular gas main sequence'' (MGMS; $\Sigma_{\mathrm{mol}}$ vs. $\Sigma_*$) as a companion to the aforementioned relationships \citep[e.g.][]{lin2019a}.
By simultaneously examining these three correlations one can gain insight into the physical mechanisms 
that 
lead to
the star formation main sequence. Here we compare 
the MGMS and $\log~\mathrm{SFR}$-$\log~M_\mathrm{mol}$ relationships of red misfit and blue active galaxies (Figure~\ref{fig:ks_mgms}). 
Each of these relationships shows strong correlations (Pearson-$r$ of detections ranging from $0.68$ to $0.86$).
We used \texttt{linmix} \citep{kelly2007} to fit lines to each of these plots, taking uncertainties in both variables and 
upper limits in $M_\mathrm{mol}$ into account.
For the $\log~\mathrm{SFR}$-$\log~M_\mathrm{mol}$  relation, the fits were done with $\log \mathrm{SFR}$ on the x-axis in order to include upper limits, and the best-fit equations were inverted to match how this relationship is usually shown with gas on the x-axis and SFR on the y-axis.

In the top row of Figure~\ref{fig:ks_mgms}, by comparing the points and fits with lines of constant $M_\mathrm{mol}/M_\star$ (dotted lines), we see that red misfits have lower molecular fractions (between 1 and 10 per cent) than blue actives (mostly around 10 per cent). The fit to blue actives is nearly linear \citep[similar to spatially resolved work e.g.,][]{lin2019a}, and the intercept is close to the value of the restricted mean from Table~\ref{tab:ks_tests} (dashed line). Red misfits however show a significantly shallower (sub-linear) slope than blue actives (a difference of $\sim 6\sigma$), causing the best-fit intercept to differ more significantly from the restricted mean from Table~\ref{tab:ks_tests} (dashed line). The difference between these populations is most striking at larger stellar masses. The shallower slope in the MGMS for red misfits  
suggests physical difference between these two populations -- the molecular gas content of red misfits is lower than that of blue actives at fixed stellar mass, but only at high stellar masses. Additionally, the correlation between $\log~M_\mathrm{mol}$ and $\log~M_\star$ is significantly weaker for red misfits ($r=0.68$) than for blue actives ($r=0.86$).

In the bottom row of Figure~\ref{fig:ks_mgms}, by comparing the data points and linear fits with lines of constant $t_\mathrm{mol}$, we see that red misfits have slightly longer $t_\mathrm{mol}$ than blue actives (echoing our earlier results). In contrast to the MGMS plots, the slope of the red misfit and blue active $\log~\mathrm{SFR}$-$\log~M_\mathrm{mol}$  relations are 
not significantly different (agree within $\sim 1\sigma$). The $\log~\mathrm{SFR}$-$\log~M_\mathrm{mol}$  slopes of red misfits and blue actives are both slightly super-linear. 
This near-linearity results in the best-fit intercepts being close to the restricted means from Table~\ref{tab:ks_tests} (dashed lines). The similarity of the gas depletion time relationships, combined with 
the MGMS results, and the fact that red misfits tend to lie below the SFMS (e.g., Figure~\ref{fig:sfms_gas}) suggests that red misfits have lower than average star formation rates due to a lack of molecular gas rather than
inefficient star formation.

\subsection{Impact of $M_\star$ differences between red misfits and blue actives on cold gas scaling relations}\label{sec:mstar_bias}

The $M_\mathrm{mol}/M_\star$ (and $M_\mathrm{gas}/M_\star$) scaling relations for red misfits and blue 
actives come closer together when we restrict the stellar mass range (Section~\ref{sec:scaling_relations}). We investigate how much of the residual difference between the red and blue points in the bottom row of Figure~\ref{fig:fgas_delta_ms} can be explained by differences in the median stellar mass of blue actives and red misfits within each bin.

We correct the $M_\mathrm{mol}/M_\star$ measurements of red misfits in a given $\Delta$ MS bin as follows. 
First we assume that the two following empirical relationships for red misfits hold:
\begin{equation}\label{eq:rm1}
\log~\mathrm{SFR} = a_1 + b_1 \log~M_\star,
\end{equation}
where $a_1 = -5.25$ and $b_1 = 0.48$ (middle panel of Figure~\ref{fig:sfms_fits_rm_ba}), and
\begin{equation}\label{eq:rm2}
\log~\mathrm{SFR} = a_2 + b_2 \log~M_\mathrm{mol},
\end{equation}
where $a_2 = -10.22$ and $b_2 = 1.12$ (bottom right panel of Figure~\ref{fig:ks_mgms}).

Consider a red misfit galaxy with stellar mass $\log M_{\star,\mathrm{RM}}$, molecular gas mass $\log M_{\mathrm{mol,~RM}}$ lying in $\Delta \mathrm{MS}$ bin $i$. Let  
\begin{equation}
\Delta \log M_\star \equiv \log M_{\star,\mathrm{RM}} - \mathrm{med}(\log M_\star)_{\mathrm{BA},i},
\end{equation} where 
$\mathrm{med}(\log M_\star)_{\mathrm{BA},i}$ is the median $\log~M_\star$ of blue actives in bin $\Delta \mathrm{MS}$ bin $i$.
We can estimate the molecular gas mass that this red misfit would have if its stellar mass 
was equal to $\mathrm{med}(\log M_\star)_{\mathrm{BA},i}$, by plugging the change in $\log \mathrm{SFR}$ 
in Equation~\ref{eq:rm1} corresponding to $\Delta \log M_\star$ into Equation~\ref{eq:rm2}, and solving for the change in molecular gas mass
\begin{equation}
\Delta\log~M_\mathrm{mol} = \frac{b_1}{b_2}~\Delta\log M_\star.
\end{equation}
The corrected $\log~M_\mathrm{mol}$ is then
\begin{equation}
\log~M_\mathrm{mol,~RM,~corr.} = \log M_{\mathrm{mol,~RM}} + \frac{b_1}{b_2}~\Delta\log M_\star.
\end{equation}
We correct the $\log~M_\mathrm{mol}/M_\star$ measurements of each red misfit in this way, and recompute the restricted 
mean $\log~M_\mathrm{mol}/M_\star$, shown as the purple stars in Figure~\ref{fig:fgas_delta_ms}.
This correction for stellar mass differences brings red misfit and blue actives even closer together (bottom left panel of Figure~\ref{fig:fgas_delta_ms}).

For the total gas scaling relations, we correct the molecular gas masses as outlined above, and we also correct the \hone\ mass fractions such that the full sample relation between $\log~M_\mathrm{HI}$ and $\log~M_\star$ from \citet{brown2015} holds (see their Figures 4 and 5, where they find a slope of $-0.85$ for this relation). The corrected red misfit $\log~M_\mathrm{gas}/M_\star$ points are shown in the bottom right panel of Figure~\ref{fig:fgas_delta_ms}. This correction brings the red misfit and blue active $\log~M_\mathrm{gas}/M_\star$ scaling relations even closer together, but red misfits still have lower fractions around $\Delta~\mathrm{MS}\sim 0$ dex.

\section{Discussion}
\label{sec:discussion}

Our findings show that red misfits and blue actives have different molecular and total gas mass fractions,  different dust mass fractions, and slightly different molecular gas depletion times, but similar total gas depletion times, and similar dust-to-gas ratios. We showed that red misfits have lower $M_\mathrm{mol}/M_\star$ and $M_\mathrm{gas}/M_\star$ ratios than blue actives on average (Section~\ref{sec:char}) and
as functions of offset from the main sequence (Section~\ref{sec:scaling_relations}). 
We showed that red misfits 
have a significantly shallower slope than blue actives in the molecular gas main sequence (Section~\ref{sec:ks_mgms}), and that red misfits and blue actives have consistent $\log~\mathrm{SFR}$-$\log~M_\mathrm{mol}$ relations (Section~\ref{sec:ks_mgms}).

We found that the dust content of red misfits is similar (based on the DGR) or lower than (based on $M_\mathrm{dust}/M_\star$) that of blue actives, which supports the claims from \citet{evans2018} that the red colours of red misfits are not due to dust reddening. Their red colours are therefore likely due to the presence of old stellar populations. However, $g-r$ colour is not as sensitive to young stellar populations as $u-r$ or $\mathrm{NUV}-r$, and so a red $g-r$ colour does not necessarily indicate a red $u-r$ or $\mathrm{NUV}-r$ colour. Indeed, by definition, red misfits are actively forming stars, and so they must host young stellar populations.

We find that red misfits have lower molecular gas fractions, and even lower total gas fractions, than blue actives, while $t_\mathrm{mol}$ and $t_\mathrm{gas}$ of red misfits and blue actives follow similar relationships.
Red misfits tend to lie on or below the main sequence while blue actives tend to lie on or above the main sequence. 
After correcting for different median stellar masses between red misfits and blue actives, their  $M_\mathrm{mol}/M_\star$ and $M_\mathrm{gas}/M_\star$ scaling relations become more similar. However, red misfits still tend to have lower total gas content particularly on the main sequence. Taken together, these results suggest that the lower star formation rates of 
red misfits lying on or near the main sequence are due to bottlenecks in the gas supply rather than reduced star formation efficiency. Our findings that the difference in total gas mass fraction is larger than that of molecular gas mass fraction suggests that the long-term fuel for star formation has been depleted. The fact that the molecular gas mass fraction of red misfits is lowest compared to blue actives at high stellar masses suggests that red misfits have depleted their gas supply by forming stars and are on their way toward the red sequence. 
Taking all of our findings together with those of \citet{evans2018}, we suggest that red misfits are 
not a single class of galaxies, but rather a mix of galaxies in different states whose behaviour depends on position relative to the SFMS. However, when we narrow in on galaxies on or slightly below the main sequence, and control for stellar mass biases, red misfits have lower total gas content than blue actives.

One limitation of the present work is that we combined several datasets 
together, and so our sample has a complex selection function. Another limitation is that
we only used 850 \um\ fluxes to estimate dust masses; a more optimal method would be to use 
infrared-to-submillimeter SED fitting. Unfortunately, our JCMT Semester 18B SCUBA-2 targets were
selected from the xCOLD GASS sample and this sample
does not overlap significantly with H-ATLAS and so the required infrared data do not exist like they 
do for JINGLE galaxies. In the interest of using the same method for all galaxies with SCUBA-2 data, 
we used the \citet{lamperti2019} scaling relations to estimate a dust temperature and spectral index for each galaxy.

\section{Conclusions}

By analyzing trends of molecular and total cold gas mass fractions and depletion times, 
we have found that red misfit and blue active galaxies do not show strong differences in depletion times,
but their gas mass fractions are significantly different, and they exhibit significantly different scaling relations with offset from the main sequence and stellar mass. This suggests that red misfits 
are more limited than blue actives in both their near term and long term gas supply rather than the rate with which they 
are turning the gas into stars.
This is also likely due to the fact that red misfits below the main sequence tend to be more massive than 
blue actives. Thus red misfits have about the same amount of gas but are more massive. 
We also found that the dust-to-stellar ratios of red misfits are lower than that of blue actives, while their dust-to-gas ratios follow similar distributions. 

Our results suggest that by selecting galaxies based on optical $g-r$ colour and specific star formation rate 
simultaneously, high mass galaxies that are classified as red and star forming (red misfits) are  actively quenching after depleting their gas supply through star formation, while red star-forming galaxies with low stellar masses either had limited gas supply to begin with or had their gas removed prematurely (e.g. due to environmental effects such as ram pressure stripping). %

\section*{Acknowledgements}

We thank the anonymous referee for their comments which helped to improve the manuscript.
The James Clerk Maxwell Telescope is operated by the East Asian Observatory on behalf of The National Astronomical Observatory of Japan; Academia Sinica Institute of Astronomy and Astrophysics; the Korea Astronomy and Space Science Institute; Center for Astronomical Mega-Science (as well as the National Key R\&D Program of China with No. 2017YFA0402700). Additional funding support is provided by the Science and Technology Facilities Council of the United Kingdom and participating universities and organizations in the United Kingdom and Canada.
Additional funds for the construction of SCUBA-2 were provided by the Canada Foundation for Innovation.
The authors wish to recognize and acknowledge the very significant cultural role and reverence that the summit of Maunakea has always had within the Indigenous Hawaiian community.  We are most fortunate to have the opportunity to conduct observations from this mountain.

H. S. H. acknowledges the support by the National Research Foundation of Korea (NRF) grant funded by the Korea government (MSIT) (No. 2021R1A2C1094577).
Y. G. acknowledges funding from the National Natural Science Foundation of China (NSFC, No. 12033004).
M. T. S. acknowledges support from a Scientific Exchanges visitor fellowship (IZSEZO\_202357) from the Swiss National Science Foundation.
T. X. acknowledges support from the National Natural Science Foundation of China (grant No. 11973030).
L. C. P. and C. D. W. acknowledge support from the Natural Science and Engineering Research Council of Canada and C. D. W. acknowledges support from the Canada Research Chairs program.

\section{Data Availability}

The data underlying this article are available in the article (Table~\ref{tab:jcmt_18a}, Table~\ref{tab:scuba2_18b}), public domain sources, and a private catalog (JINGLE).

\bibliographystyle{mnras}
\bibliography{wise_co_paper_v2} %

\begin{thebibliography}{}
\makeatletter
\relax
\def\mn@urlcharsother{\let\do\@makeother \do\$\do\&\do\#\do\^\do\_\do\%\do\~}
\def\mn@doi{\begingroup\mn@urlcharsother \@ifnextchar [ {\mn@doi@}
  {\mn@doi@[]}}
\def\mn@doi@[#1]#2{\def\@tempa{#1}\ifx\@tempa\@empty \href
  {http://dx.doi.org/#2} {doi:#2}\else \href {http://dx.doi.org/#2} {#1}\fi
  \endgroup}
\def\mn@eprint#1#2{\mn@eprint@#1:#2::\@nil}
\def\mn@eprint@arXiv#1{\href {http://arxiv.org/abs/#1} {{\tt arXiv:#1}}}
\def\mn@eprint@dblp#1{\href {http://dblp.uni-trier.de/rec/bibtex/#1.xml}
  {dblp:#1}}
\def\mn@eprint@#1:#2:#3:#4\@nil{\def\@tempa {#1}\def\@tempb {#2}\def\@tempc
  {#3}\ifx \@tempc \@empty \let \@tempc \@tempb \let \@tempb \@tempa \fi \ifx
  \@tempb \@empty \def\@tempb {arXiv}\fi \@ifundefined
  {mn@eprint@\@tempb}{\@tempb:\@tempc}{\expandafter \expandafter \csname
  mn@eprint@\@tempb\endcsname \expandafter{\@tempc}}}

\bibitem[\protect\citeauthoryear{{Astropy Collaboration} et~al.,}{{Astropy
  Collaboration} et~al.}{2018}]{astropy-collaboration2018}
{Astropy Collaboration} et~al., 2018, \mn@doi [\aj] {10.3847/1538-3881/aabc4f},
  \href {https://ui.adsabs.harvard.edu/abs/2018AJ....156..123A} {156, 123}

\bibitem[\protect\citeauthoryear{{Belfiore} et~al.,}{{Belfiore}
  et~al.}{2017}]{belfiore2017}
{Belfiore} F.,  et~al., 2017, \mn@doi [\mnras] {10.1093/mnras/stw3211}, \href
  {http://adsabs.harvard.edu/abs/2017MNRAS.466.2570B} {466, 2570}

\bibitem[\protect\citeauthoryear{{Bigiel}, {Leroy}, {Walter}, {Brinks}, {de
  Blok}, {Madore}  \& {Thornley}}{{Bigiel} et~al.}{2008}]{bigiel2008}
{Bigiel} F.,  {Leroy} A.,  {Walter} F.,  {Brinks} E.,  {de Blok} W.~J.~G.,
  {Madore} B.,   {Thornley} M.~D.,  2008, \mn@doi [\aj]
  {10.1088/0004-6256/136/6/2846}, \href
  {http://adsabs.harvard.edu/abs/2008AJ....136.2846B} {136, 2846}

\bibitem[\protect\citeauthoryear{{Bolatto}, {Wolfire}  \& {Leroy}}{{Bolatto}
  et~al.}{2013}]{bolatto2013}
{Bolatto} A.~D.,  {Wolfire} M.,   {Leroy} A.~K.,  2013, \mn@doi [\araa]
  {10.1146/annurev-astro-082812-140944}, \href
  {http://adsabs.harvard.edu/abs/2013ARA%26A..51..207B} {51, 207}

\bibitem[\protect\citeauthoryear{{Brown}, {Catinella}, {Cortese}, {Kilborn},
  {Haynes}  \& {Giovanelli}}{{Brown} et~al.}{2015}]{brown2015}
{Brown} T.,  {Catinella} B.,  {Cortese} L.,  {Kilborn} V.,  {Haynes} M.~P.,
  {Giovanelli} R.,  2015, \mn@doi [\mnras] {10.1093/mnras/stv1311}, \href
  {https://ui.adsabs.harvard.edu/abs/2015MNRAS.452.2479B} {452, 2479}

\bibitem[\protect\citeauthoryear{{Brownson}, {Belfiore}, {Maiolino}, {Lin}  \&
  {Carniani}}{{Brownson} et~al.}{2020}]{brownson2020}
{Brownson} S.,  {Belfiore} F.,  {Maiolino} R.,  {Lin} L.,   {Carniani} S.,
  2020, \mn@doi [\mnras] {10.1093/mnrasl/slaa128}, \href
  {https://ui.adsabs.harvard.edu/abs/2020MNRAS.498L..66B} {498, L66}

\bibitem[\protect\citeauthoryear{{Catinella} et~al.,}{{Catinella}
  et~al.}{2018}]{catinella2018}
{Catinella} B.,  et~al., 2018, \mn@doi [\mnras] {10.1093/mnras/sty089}, \href
  {https://ui.adsabs.harvard.edu/abs/2018MNRAS.476..875C} {476, 875}

\bibitem[\protect\citeauthoryear{{Chown}, {Li}, {Parker}, {Wilson}, {Li}  \&
  {Gao}}{{Chown} et~al.}{2021}]{chown2021}
{Chown} R.,  {Li} C.,  {Parker} L.,  {Wilson} C.~D.,  {Li} N.,   {Gao} Y.,
  2021, \mn@doi [\mnras] {10.1093/mnras/staa3288}, \href
  {https://ui.adsabs.harvard.edu/abs/2021MNRAS.500.1261C} {500, 1261}

\bibitem[\protect\citeauthoryear{{Clark}, {Schofield}, {Gomez}  \&
  {Davies}}{{Clark} et~al.}{2016}]{clark2016}
{Clark} C. J.~R.,  {Schofield} S.~P.,  {Gomez} H.~L.,   {Davies} J.~I.,  2016,
  \mn@doi [\mnras] {10.1093/mnras/stw647}, \href
  {https://ui.adsabs.harvard.edu/abs/2016MNRAS.459.1646C} {459, 1646}

\bibitem[\protect\citeauthoryear{{Coenda}, {Mart{\'\i}nez}  \&
  {Muriel}}{{Coenda} et~al.}{2018}]{coenda2018}
{Coenda} V.,  {Mart{\'\i}nez} H.~J.,   {Muriel} H.,  2018, \mn@doi [\mnras]
  {10.1093/mnras/stx2707}, \href
  {https://ui.adsabs.harvard.edu/abs/2018MNRAS.473.5617C} {473, 5617}

\bibitem[\protect\citeauthoryear{{Colombo} et~al.,}{{Colombo}
  et~al.}{2020}]{colombo2020}
{Colombo} D.,  et~al., 2020, \mn@doi [\aap] {10.1051/0004-6361/202039005},
  \href {https://ui.adsabs.harvard.edu/abs/2020A&A...644A..97C} {644, A97}

\bibitem[\protect\citeauthoryear{{Eales} et~al.,}{{Eales}
  et~al.}{2018}]{eales2018}
{Eales} S.~A.,  et~al., 2018, \mn@doi [\mnras] {10.1093/mnras/sty2220}, \href
  {https://ui.adsabs.harvard.edu/abs/2018MNRAS.481.1183E} {481, 1183}

\bibitem[\protect\citeauthoryear{{Ellison} et~al.,}{{Ellison}
  et~al.}{2020}]{ellison2020}
{Ellison} S.~L.,  et~al., 2020, \mn@doi [\mnras] {10.1093/mnrasl/slz179}, \href
  {https://ui.adsabs.harvard.edu/abs/2020MNRAS.493L..39E} {493, L39}

\bibitem[\protect\citeauthoryear{{Evans}, {Parker}  \& {Roberts}}{{Evans}
  et~al.}{2018}]{evans2018}
{Evans} F.~A.,  {Parker} L.~C.,   {Roberts} I.~D.,  2018, \mn@doi [\mnras]
  {10.1093/mnras/sty581}, \href
  {https://ui.adsabs.harvard.edu/abs/2018MNRAS.476.5284E} {476, 5284}

\bibitem[\protect\citeauthoryear{{Feldmann}}{{Feldmann}}{2019}]{feldmann2019}
{Feldmann} R.,  2019, \mn@doi [Astronomy and Computing]
  {10.1016/j.ascom.2019.100331}, \href
  {https://ui.adsabs.harvard.edu/abs/2019A&C....2900331F} {29, 100331}

\bibitem[\protect\citeauthoryear{{Feldmann}}{{Feldmann}}{2020}]{feldmann2020}
{Feldmann} R.,  2020, \mn@doi [Communications Physics]
  {10.1038/s42005-020-00493-0}, \href
  {https://ui.adsabs.harvard.edu/abs/2020CmPhy...3..226F} {3, 226}

\bibitem[\protect\citeauthoryear{{Haynes} et~al.,}{{Haynes}
  et~al.}{2018}]{haynes2018}
{Haynes} M.~P.,  et~al., 2018, \mn@doi [\apj] {10.3847/1538-4357/aac956}, \href
  {https://ui.adsabs.harvard.edu/abs/2018ApJ...861...49H} {861, 49}

\bibitem[\protect\citeauthoryear{{Kelly}}{{Kelly}}{2007}]{kelly2007}
{Kelly} B.~C.,  2007, \mn@doi [\apj] {10.1086/519947}, \href
  {https://ui.adsabs.harvard.edu/abs/2007ApJ...665.1489K} {665, 1489}

\bibitem[\protect\citeauthoryear{{Kennicutt}}{{Kennicutt}}{1989}]{kennicutt1989}
{Kennicutt} Robert~C. J.,  1989, \mn@doi [\apj] {10.1086/167834}, \href
  {https://ui.adsabs.harvard.edu/abs/1989ApJ...344..685K} {344, 685}

\bibitem[\protect\citeauthoryear{{Kennicutt} Robert~C. et~al.,}{{Kennicutt}
  et~al.}{2007}]{kennicutt2007}
{Kennicutt} Robert~C. J.,  et~al., 2007, \mn@doi [\apj] {10.1086/522300}, \href
  {https://ui.adsabs.harvard.edu/abs/2007ApJ...671..333K} {671, 333}

\bibitem[\protect\citeauthoryear{{Lamperti} et~al.,}{{Lamperti}
  et~al.}{2019}]{lamperti2019}
{Lamperti} I.,  et~al., 2019, \mn@doi [\mnras] {10.1093/mnras/stz2311}, \href
  {https://ui.adsabs.harvard.edu/abs/2019MNRAS.489.4389L} {489, 4389}

\bibitem[\protect\citeauthoryear{{Leroy}, {Walter}, {Brinks}, {Bigiel}, {de
  Blok}, {Madore}  \& {Thornley}}{{Leroy} et~al.}{2008}]{leroy2008}
{Leroy} A.~K.,  {Walter} F.,  {Brinks} E.,  {Bigiel} F.,  {de Blok} W.~J.~G.,
  {Madore} B.,   {Thornley} M.~D.,  2008, \mn@doi [\aj]
  {10.1088/0004-6256/136/6/2782}, \href
  {http://adsabs.harvard.edu/abs/2008AJ....136.2782L} {136, 2782}

\bibitem[\protect\citeauthoryear{{Leroy} et~al.,}{{Leroy}
  et~al.}{2013}]{leroy2013}
{Leroy} A.~K.,  et~al., 2013, \mn@doi [\aj] {10.1088/0004-6256/146/2/19}, \href
  {http://adsabs.harvard.edu/abs/2013AJ....146...19L} {146, 19}

\bibitem[\protect\citeauthoryear{{Li} et~al.,}{{Li} et~al.}{2015}]{li2015}
{Li} C.,  et~al., 2015, \mn@doi [\apj] {10.1088/0004-637X/804/2/125}, \href
  {http://adsabs.harvard.edu/abs/2015ApJ...804..125L} {804, 125}

\bibitem[\protect\citeauthoryear{{Lilly}, {Carollo}, {Pipino}, {Renzini}  \&
  {Peng}}{{Lilly} et~al.}{2013}]{lilly2013}
{Lilly} S.~J.,  {Carollo} C.~M.,  {Pipino} A.,  {Renzini} A.,   {Peng} Y.,
  2013, \mn@doi [\apj] {10.1088/0004-637X/772/2/119}, \href
  {http://adsabs.harvard.edu/abs/2013ApJ...772..119L} {772, 119}

\bibitem[\protect\citeauthoryear{{Lin} et~al.,}{{Lin} et~al.}{2017}]{lin2017}
{Lin} L.,  et~al., 2017, \mn@doi [\apj] {10.3847/1538-4357/aa96ae}, \href
  {http://adsabs.harvard.edu/abs/2017ApJ...851...18L} {851, 18}

\bibitem[\protect\citeauthoryear{{Lin} et~al.,}{{Lin} et~al.}{2019}]{lin2019a}
{Lin} L.,  et~al., 2019, \mn@doi [\apjl] {10.3847/2041-8213/ab4815}, \href
  {https://ui.adsabs.harvard.edu/abs/2019ApJ...884L..33L} {884, L33}

\bibitem[\protect\citeauthoryear{{Lin} et~al.,}{{Lin} et~al.}{2020}]{lin2020a}
{Lin} L.,  et~al., 2020, \mn@doi [\apj] {10.3847/1538-4357/abba3a}, \href
  {https://ui.adsabs.harvard.edu/abs/2020ApJ...903..145L} {903, 145}

\bibitem[\protect\citeauthoryear{{Lin} et~al.,}{{Lin} et~al.}{2022}]{lin2022}
{Lin} L.,  et~al., 2022, \mn@doi [\apj] {10.3847/1538-4357/ac4ccc}, \href
  {https://ui.adsabs.harvard.edu/abs/2022ApJ...926..175L} {926, 175}

\bibitem[\protect\citeauthoryear{{Mancini} et~al.,}{{Mancini}
  et~al.}{2019}]{mancini2019}
{Mancini} C.,  et~al., 2019, \mn@doi [\mnras] {10.1093/mnras/stz2130}, \href
  {https://ui.adsabs.harvard.edu/abs/2019MNRAS.489.1265M} {489, 1265}

\bibitem[\protect\citeauthoryear{{Mok} et~al.,}{{Mok} et~al.}{2016}]{mok2016}
{Mok} A.,  et~al., 2016, \mn@doi [\mnras] {10.1093/mnras/stv2958}, \href
  {https://ui.adsabs.harvard.edu/abs/2016MNRAS.456.4384M} {456, 4384}

\bibitem[\protect\citeauthoryear{{Pettini} \& {Pagel}}{{Pettini} \&
  {Pagel}}{2004}]{pettini2004}
{Pettini} M.,  {Pagel} B. E.~J.,  2004, \mn@doi [\mnras]
  {10.1111/j.1365-2966.2004.07591.x}, \href
  {https://ui.adsabs.harvard.edu/abs/2004MNRAS.348L..59P} {348, L59}

\bibitem[\protect\citeauthoryear{{Popesso} et~al.,}{{Popesso}
  et~al.}{2019}]{popesso2019}
{Popesso} P.,  et~al., 2019, \mn@doi [\mnras] {10.1093/mnras/sty3210}, \href
  {https://ui.adsabs.harvard.edu/abs/2019MNRAS.483.3213P} {483, 3213}

\bibitem[\protect\citeauthoryear{{Saintonge} \& {Catinella}}{{Saintonge} \&
  {Catinella}}{2022}]{saintonge2022}
{Saintonge} A.,  {Catinella} B.,  2022, arXiv e-prints, \href
  {https://ui.adsabs.harvard.edu/abs/2022arXiv220200690S} {p. arXiv:2202.00690}

\bibitem[\protect\citeauthoryear{{Saintonge} et~al.,}{{Saintonge}
  et~al.}{2011}]{saintonge2011}
{Saintonge} A.,  et~al., 2011, \mn@doi [\mnras]
  {10.1111/j.1365-2966.2011.18677.x}, \href
  {http://adsabs.harvard.edu/abs/2011MNRAS.415...32S} {415, 32}

\bibitem[\protect\citeauthoryear{{Saintonge} et~al.,}{{Saintonge}
  et~al.}{2016}]{saintonge2016}
{Saintonge} A.,  et~al., 2016, \mn@doi [\mnras] {10.1093/mnras/stw1715}, \href
  {https://ui.adsabs.harvard.edu/abs/2016MNRAS.462.1749S} {462, 1749}

\bibitem[\protect\citeauthoryear{{Saintonge} et~al.,}{{Saintonge}
  et~al.}{2017}]{saintonge2017}
{Saintonge} A.,  et~al., 2017, \mn@doi [\apjs] {10.3847/1538-4365/aa97e0},
  \href {http://adsabs.harvard.edu/abs/2017ApJS..233...22S} {233, 22}

\bibitem[\protect\citeauthoryear{{Saintonge} et~al.,}{{Saintonge}
  et~al.}{2018}]{saintonge2018}
{Saintonge} A.,  et~al., 2018, \mn@doi [\mnras] {10.1093/mnras/sty2499}, \href
  {https://ui.adsabs.harvard.edu/abs/2018MNRAS.481.3497S} {481, 3497}

\bibitem[\protect\citeauthoryear{{Salim}}{{Salim}}{2014}]{salim2014}
{Salim} S.,  2014, \mn@doi [Serbian Astronomical Journal]
  {10.2298/SAJ1489001S}, \href
  {https://ui.adsabs.harvard.edu/abs/2014SerAJ.189....1S} {189, 1}

\bibitem[\protect\citeauthoryear{{Salim}, {Boquien}  \& {Lee}}{{Salim}
  et~al.}{2018}]{salim2018}
{Salim} S.,  {Boquien} M.,   {Lee} J.~C.,  2018, \mn@doi [\apj]
  {10.3847/1538-4357/aabf3c}, \href
  {https://ui.adsabs.harvard.edu/abs/2018ApJ...859...11S} {859, 11}

\bibitem[\protect\citeauthoryear{{S{\'a}nchez} et~al.,}{{S{\'a}nchez}
  et~al.}{2021}]{sanchez2021}
{S{\'a}nchez} S.~F.,  et~al., 2021, \mn@doi [\mnras] {10.1093/mnras/stab442},
  \href {https://ui.adsabs.harvard.edu/abs/2021MNRAS.503.1615S} {503, 1615}

\bibitem[\protect\citeauthoryear{{Sargent} et~al.,}{{Sargent}
  et~al.}{2014}]{sargent2014}
{Sargent} M.~T.,  et~al., 2014, \mn@doi [\apj] {10.1088/0004-637X/793/1/19},
  \href {https://ui.adsabs.harvard.edu/abs/2014ApJ...793...19S} {793, 19}

\bibitem[\protect\citeauthoryear{{Schawinski} et~al.,}{{Schawinski}
  et~al.}{2014}]{schawinski2014}
{Schawinski} K.,  et~al., 2014, \mn@doi [\mnras] {10.1093/mnras/stu327}, \href
  {https://ui.adsabs.harvard.edu/abs/2014MNRAS.440..889S} {440, 889}

\bibitem[\protect\citeauthoryear{{Smethurst} et~al.,}{{Smethurst}
  et~al.}{2015}]{smethurst2015}
{Smethurst} R.~J.,  et~al., 2015, \mn@doi [\mnras] {10.1093/mnras/stv161},
  \href {https://ui.adsabs.harvard.edu/abs/2015MNRAS.450..435S} {450, 435}

\bibitem[\protect\citeauthoryear{{Smith} et~al.,}{{Smith}
  et~al.}{2019}]{smith2019}
{Smith} M. W.~L.,  et~al., 2019, \mn@doi [\mnras] {10.1093/mnras/stz1102},
  \href {https://ui.adsabs.harvard.edu/abs/2019MNRAS.486.4166S} {486, 4166}

\bibitem[\protect\citeauthoryear{{Tacconi} et~al.,}{{Tacconi}
  et~al.}{2013}]{tacconi2013}
{Tacconi} L.~J.,  et~al., 2013, \mn@doi [\apj] {10.1088/0004-637X/768/1/74},
  \href {https://ui.adsabs.harvard.edu/abs/2013ApJ...768...74T} {768, 74}

\bibitem[\protect\citeauthoryear{{Tacconi} et~al.,}{{Tacconi}
  et~al.}{2018}]{tacconi2018}
{Tacconi} L.~J.,  et~al., 2018, \mn@doi [\apj] {10.3847/1538-4357/aaa4b4},
  \href {http://adsabs.harvard.edu/abs/2018ApJ...853..179T} {853, 179}

\bibitem[\protect\citeauthoryear{{Yajima} et~al.,}{{Yajima}
  et~al.}{2021}]{yajima2021}
{Yajima} Y.,  et~al., 2021, \mn@doi [\pasj] {10.1093/pasj/psaa119}, \href
  {https://ui.adsabs.harvard.edu/abs/2021PASJ...73..257Y} {73, 257}

\bibitem[\protect\citeauthoryear{{York} et~al.,}{{York}
  et~al.}{2000}]{york2000}
{York} D.~G.,  et~al., 2000, \mn@doi [\aj] {10.1086/301513}, \href
  {http://adsabs.harvard.edu/abs/2000AJ....120.1579Y} {120, 1579}

\makeatother
\end{thebibliography}

\appendix

\section{New JCMT CO(2-1) measurements of red misfits}

Here we present the CO(2-1) measurements of red misfits using JCMT (Table~\ref{tab:jcmt_18a}).

\begin{table*}
\caption{New JCMT CO(2-1) measurements of red misfits selected from the JINGLE sample.}
\label{tab:jcmt_18a}
\begin{center}
\begin{tabular}{cccccccc}
\hline
ObjID & RA (J2000) & Dec (J2000) & $z$ & $\log M_\star$ & $\log \mathrm{SFR}$ & $\log L_\mathrm{CO(1-0)}$ & $\log M_\mathrm{mol}$ \\
& deg & deg & & M$_\odot$ & M$_\odot$ yr$^{-1}$ & K km s$^{-1}$ pc$^2$ & M$_\odot$ \\
 (1) & (2) & (3) & (4) & (5) & (6) & (7) & (8) \\
  \hline
$1237665024374865968$ & $203.305$ & $33.110$ & $0.0240$ & $10.78$ & $0.25$ & $9.40\pm0.05$ & $10.04\pm0.68$ \\
$1237650762924621828$ & $173.539$ & $-1.595$ & $0.0230$ & $10.45$ & $0.29$ & $9.29\pm0.06$ & $9.93\pm0.70$ \\
$1237654604239274286$ & $131.119$ & $2.064$ & $0.0250$ & $10.32$ & $0.51$ & $9.22\pm0.07$ & $9.86\pm0.71$ \\
$1237650372092035132$ & $178.878$ & $-1.261$ & $0.0190$ & $10.08$ & $0.34$ & $8.73\pm0.12$ & $9.37\pm0.75$ \\
$1237650761854222501$ & $181.212$ & $-2.438$ & $0.0200$ & $10.21$ & $-0.08$ & $8.55\pm0.12$ & $9.19\pm0.75$ \\
$1237665126939295886$ & $201.313$ & $32.671$ & $0.0400$ & $10.82$ & $0.52$ & $<9.05$ & $<9.69$ \\
$1237671128051220863$ & $175.912$ & $-1.647$ & $0.0430$ & $10.46$ & $0.60$ & $<9.03$ & $<9.67$ \\
$1237648720695001218$ & $182.575$ & $-0.518$ & $0.0350$ & $11.02$ & $0.27$ & $<8.94$ & $<9.57$ \\
$1237648703516115092$ & $212.612$ & $-0.832$ & $0.0250$ & $10.71$ & $0.59$ & $<8.65$ & $<9.29$ \\
$1237648705663664205$ & $212.740$ & $1.036$ & $0.0250$ & $10.58$ & $0.24$ & $<8.64$ & $<9.28$ \\
$1237648722820661612$ & $132.804$ & $1.062$ & $0.0270$ & $9.94$ & $-0.10$ & $<8.55$ & $<9.19$ \\
$1237650762927308812$ & $179.693$ & $-1.466$ & $0.0210$ & $9.72$ & $0.03$ & $<8.49$ & $<9.13$ \\
\hline
\multicolumn{8}{l}{(1) SDSS photometric identification number.}\\
\multicolumn{8}{l}{(5) Stellar mass from the GSWLC-M2 or A2 catalog (if unavailable in M2).}\\
\multicolumn{8}{l}{(6) SFR from the GSWLC-M2 or A2 catalog (if unavailable in M2).}\\
\multicolumn{8}{l}{(7) Measured CO(1-0) luminosity (converted from 2-1 assuming $r_{21}=0.7$).}\\
\multicolumn{8}{l}{(8) Measured molecular gas mass assuming $\alpha_{CO} = 4.35$.}\\
\end{tabular}
\end{center}
\end{table*}

\section{New SCUBA-2 measurements of red misfits}

Here we present the SCUBA-2 850 \um\ measurements of red misfits using JCMT (Table~\ref{tab:scuba2_18b}).

\begin{table*}
\caption{New SCUBA-2 measurements of red misfits selected from the xCOLD GASS sample.}
\label{tab:scuba2_18b}
\begin{center}
\begin{tabular}{ccccccccc}
\hline
Name & $d_L$ & $T$ & $\beta$ & Det? & $r_\mathrm{ap}$ & $r_{90}$ & $S_{850~\mathrm{\mu m}}(r\leq r_\mathrm{ap})$ & $\log M_\mathrm{dust}(r\leq r_\mathrm{ap})$ \\
 & Mpc & K &  &  & arcsec & arcsec & mJy & $M_\odot$ \\
 (1) & (2) & (3) & (4) & (5) & (6) & (7) & (8) & (9) \\
  \hline
J142720.13+025018.1 & $115.53$ & $22.64$ & $1.94$ & Y & 20.23 & 15.51 & $25.37 \pm 3.95$ & $8.18\pm 0.07$ \\
J104402.21+043946.8 & $116.29$ & $22.25$ & $1.65$ & Y & 20.06 & 15.28 & $16.46 \pm 2.90$ & $7.95\pm 0.08$ \\
J101638.39+123438.5 & $138.88$ & $22.14$ & $1.86$ & Y & 28.24 & 25.07 & $25.21 \pm 4.36$ & $8.34\pm 0.07$ \\
J100530.26+054019.4 & $196.55$ & $20.16$ & $1.94$ & Y & 21.56 & 17.20 & $16.30 \pm 2.51$ & $8.53\pm 0.07$ \\
J095144.91+353719.6 & $117.99$ & $22.93$ & $1.93$ & Y & 19.50 & 14.53 & $19.72 \pm 3.98$ & $8.08\pm 0.09$ \\
J235644.47+135435.4 & $159.92$ & $22.81$ & $1.82$ & Y & 20.77 & 16.20 & $14.29 \pm 2.98$ & $8.19\pm 0.09$ \\
J105315.29+042003.1 & $184.23$ & $22.54$ & $1.91$ & Y & 15.96 & 9.27 & $19.18 \pm 3.68$ & $8.46\pm 0.08$ \\
J100216.28+191256.3 & $71.90$ & $21.88$ & $1.92$ & Y & 19.38 & 14.38 & $13.77 \pm 3.51$ & $7.52\pm 0.11$ \\
J080442.30+154632.6 & $128.22$ & $21.32$ & $1.86$ & Y & 18.62 & 13.33 & $9.98 \pm 2.94$ & $7.89\pm 0.13$ \\
J112311.63+130703.7 & $208.32$ & $22.26$ & $1.90$ & Y & 18.13 & 12.64 & $14.51 \pm 4.23$ & $8.45\pm 0.13$ \\
J094419.42+095905.1 & $44.09$ & $20.38$ & $2.03$ & N & 16.58 & 10.28 & $<9.82$ & $<7.02$ \\
J090923.67+223050.1 & $64.52$ & $21.15$ & $2.01$ & N & 18.17 & 12.69 & $<4.15$ & $<6.95$ \\
J135845.41+203942.7 & $69.95$ & $22.29$ & $2.00$ & N & 16.56 & 10.26 & $<8.18$ & $<7.28$ \\
J232326.53+152510.4 & $189.11$ & $22.87$ & $1.96$ & N & 14.47 & 6.35 & $<10.56$ & $<8.23$ \\
J151604.47+065051.4 & $162.09$ & $22.99$ & $1.93$ & N & 21.02 & 16.52 & $<10.37$ & $<8.08$ \\
J093953.62+034850.2 & $124.82$ & $21.73$ & $1.88$ & N & 19.80 & 14.94 & $<3.89$ & $<7.45$ \\
J104251.39+055135.5 & $146.88$ & $20.90$ & $1.94$ & N & 23.48 & 19.55 & $<7.69$ & $<7.92$ \\
J122006.47+100429.2 & $191.89$ & $22.13$ & $1.92$ & N & 18.93 & 13.76 & $<6.80$ & $<8.06$ \\
J152747.42+093729.6 & $136.77$ & $21.57$ & $2.02$ & N & 17.33 & 11.46 & $<6.99$ & $<7.82$ \\
J131934.30+102717.5 & $213.27$ & $21.20$ & $1.93$ & N & 22.68 & 18.58 & $<11.19$ & $<8.40$ \\
J102508.93+133605.1 & $80.82$ & $21.56$ & $1.72$ & N & 19.07 & 13.96 & $<7.17$ & $<7.30$ \\
J142846.66+271502.4 & $63.83$ & $24.29$ & $1.90$ & N & 16.06 & 9.42 & $<7.62$ & $<7.09$ \\
J021219.38+133645.6 & $182.82$ & $22.01$ & $1.89$ & N & 16.66 & 10.42 & $<4.80$ & $<7.87$ \\
J130035.67+273427.2 & $73.74$ & $21.76$ & $1.81$ & N & 23.86 & 20.01 & $<6.54$ & $<7.20$ \\
J001947.33+003526.7 & $76.84$ & $22.52$ & $1.70$ & N & 17.88 & 12.28 & $<6.84$ & $<7.21$ \\
J020359.14+141837.3 & $189.02$ & $23.00$ & $1.91$ & N & 22.37 & 18.20 & $<12.11$ & $<8.28$ \\
J130525.44+035929.7 & $193.31$ & $22.13$ & $1.87$ & N & 15.77 & 8.93 & $<4.17$ & $<7.85$ \\
J095439.45+092640.7 & $151.93$ & $21.50$ & $1.97$ & N & 18.41 & 13.04 & $<6.71$ & $<7.88$ \\
J111738.91+263506.0 & $210.84$ & $22.38$ & $2.07$ & N & 13.61 & 4.03 & $<7.26$ & $<8.20$ \\
J231816.95+133426.6 & $174.37$ & $22.52$ & $1.84$ & N & 15.15 & 7.78 & $<5.19$ & $<7.83$ \\
J011716.09+143720.5 & $167.65$ & $19.62$ & $1.98$ & N & 15.68 & 8.76 & $<4.66$ & $<7.87$ \\
J150926.10+101718.3 & $120.84$ & $21.84$ & $1.78$ & N & 20.59 & 15.96 & $<9.99$ & $<7.81$ \\
J150204.10+064922.9 & $204.97$ & $21.47$ & $1.75$ & N & 16.90 & 10.79 & $<5.99$ & $<8.04$ \\
\hline
\multicolumn{9}{l}{(1) SDSS ID as shown in the xCOLD GASS catalog.}\\
\multicolumn{9}{l}{(2) Luminosity distance.}\\
\multicolumn{9}{l}{(3) Dust temperature estimated using Equation~\ref{eq:lamperti_t}.}\\
\multicolumn{9}{l}{(4) Modified blackbody spectral index estimated using Equation~\ref{eq:lamperti_beta}.}\\
\multicolumn{9}{l}{(5) Flag for whether this galaxy is classified as a detection or not.}\\
\multicolumn{9}{l}{(6) Aperture radius over which the 850 \micron\ flux density was measured.}\\
\multicolumn{9}{l}{(7) SDSS r-band 90 per cent Petrosian radius.}\\
\multicolumn{9}{l}{(8) 850 \micron\ flux density within $r_\mathrm{ap}$.}\\
\multicolumn{9}{l}{(9) Dust mass within $r_\mathrm{ap}$ computed using Equation~\ref{eq:mdust}.}\\
\end{tabular}
\end{center}
\end{table*}

\section{$\log~\mathrm{SFR}$ vs. $\log~\mathrm{M_\star}$ for red misfits and blue actives}

\begin{figure*}
\centering
	\includegraphics[width=\linewidth]{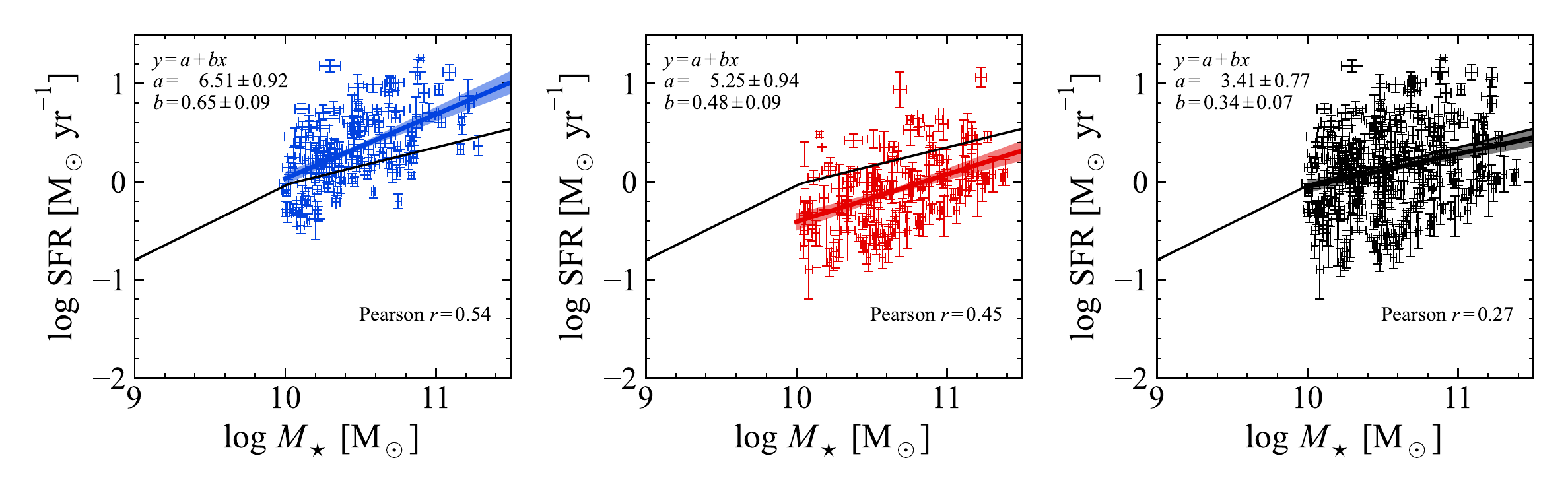}
	\caption{Linear fits to $\log~\mathrm{SFR}$ vs. $\log~\mathrm{M_\star}$ for blue actives (left), red misfits (middle) and both populations combined (right). The fits were performed using LinMix, taking uncertainties into account. The solid black lines in each panel are 
the best fit relation from \citet{popesso2019}.
    }
    \label{fig:sfms_fits_rm_ba}
\end{figure*}

Here we present fits $\log~\mathrm{SFR}$ vs. $\log~\mathrm{M_\star}$ for red misfits and blue actives, and the whole population (Figure~\ref{fig:sfms_fits_rm_ba}). The solid black lines in each panel are 
the best fit relation from \citet{popesso2019}. Blue actives and red misfits on their own do not follow the \citet{popesso2019} relation, but the fit to both populations combined is consistent with \citet{popesso2019}. The slope of the fit to red misfits is used to correct gas mass fractions in Section~\ref{sec:mstar_bias}.

\bsp	%
\label{lastpage}
\end{document}